\documentclass[epj]{svjour}
\usepackage{graphics}
\usepackage{epsfig}
\usepackage{comment}
\newcommand{\bm}[1]{ \mbox{\boldmath $#1$}  }

\begin{document}

\title{ Three-body structure of low-lying $^{18}$Ne states}

\author{J.A. Lay\inst{1} \and D.V. Fedorov\inst{2} \and A.S.~Jensen\inst{2} \and E. Garrido\inst{3} \and C. Romero-Redondo\inst{3}}
\institute{Departamento de F\'isica At\'omica, Molecular y Nuclear,
Universidad de Sevilla, Aptdo 1065, ES-41080 Sevilla, Spain  \and  Department of Physics and Astronomy,
        Aarhus University, DK-8000 Aarhus C, Denmark \and Instituto de Estructura de la Materia, CSIC, 
Serrano 123, E-28006 Madrid, Spain}

\date{\today}

\abstract{
We investigate to what extent $^{18}$Ne can be descibed as a
three-body system made of an inert $^{16}$O-core and two protons.  We
compare to experimental data and occasionally to shell model results.
We obtain three-body wave functions with the hyperspherical adiabatic
expansion method. We study the spectrum of $^{18}$Ne, the structure of
the different states and the predominant transition strengths. Two
$0^+$, two $2^+$, and one $4^+$ bound states are found where they are
all known experimentally. Also one $3^+$ close to threshold is found
and several negative parity states, $1^-$, $3^-$, $0^-$, $2^-$, most
of them bound with respect to the $^{16}$O excited $3^-$ state. The
structures are extracted as partial wave components, as spatial sizes
of matter and charge, and as probability distributions.
Electromagnetic decay rates are calculated for these states.  The
dominating decay mode for the bound states is $E2$ and occasionally
also $M1$.}


\maketitle

\section{Introduction}

Nuclear cluster structures appear in various disguises especially in
light nuclei. The cluster constituents are often nucleons and
$\alpha$-particles, possibly combined with a core-nucleus. These
structures, which appear both as ground and excited (resonance)
states, are sometimes well described as three-body systems.  The
conventional wisdom is that prominent clusters are most likely to
appear close to the threshold energy for fragmentation into the
cluster constituents.  This implies that cluster structures for
ordinary bound nuclei are more likely to appear in excited states than
in ground states, except for dripline nuclei where the ground state is
close to the nucleon threshold and dominating one or two-nucleon
structures appear \cite{jen04,tho04,fre07,bla08}.

Well-known three-body examples are the first $0^+$ resonance in
$^{12}$C \cite{alv07}, the lowest $0^+$ and $2^+$ states in $^{6}$He,
$^{6}$Be and $^{6}$Li \cite{gar06}, the $3/2^-$ ground state in
$^{11}$Li \cite{gar01}, three excited bound states in $^{12}$Be
\cite{rom08}, the ground state and four resonances in $^{17}$Ne
\cite{gar03}, the ground state and several resonances in $^{9}$Be
\cite{alv08}, and three resonances in $^{5}$H \cite{die07}.
Other nuclear states have significant admixtures of non-cluster
structure ($^{12}$C($2^{\pm}$)) \cite{alv08a} while some states are far
better described without any cluster structure ($^{12}$C($1^{\pm}$))
\cite{alv07}.  One line of investigation is to carry out the
three-body computation for a given system and compare the computed
observables with known measured values and then predict others. 

If the computed bulk structure of a nuclear state matches measurements
the description is an immediate success.  However, even for cases
where no traces of any three-body structure can be found the
computation can be considered a necessary ingredient to describe the
three-body decay of an underlying many-body resonance, examples are
$^{12}$C($1^{\pm})$ \cite{alv08b}.  Resonances decaying into three
clusters are now investigated accurately in details in complete
kinematics \cite{fyn04}. Both structure and dynamic evolution from
small to large distances are important in a description of the
momentum distributions of the decay products.  A few such decaying
structures have been investigated theoretically and compared to
available data \cite{gar06b,alv08}.

\begin{figure}
\vspace*{-0.8cm}
\epsfig{file=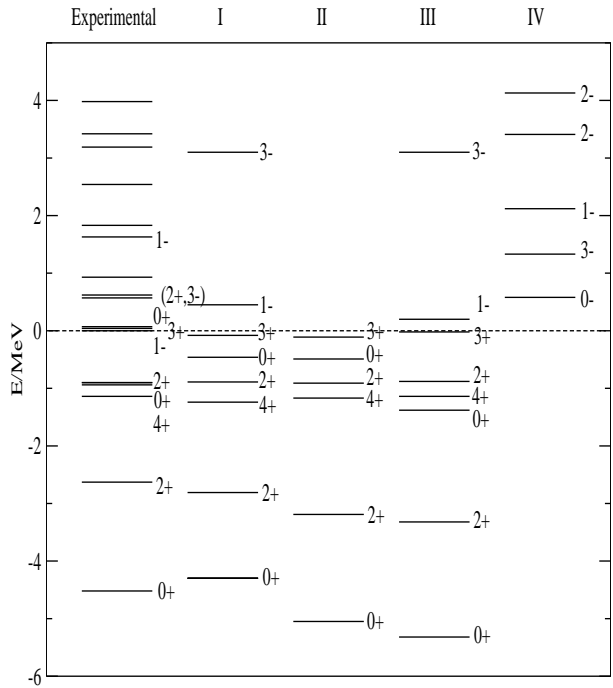,height=9.6cm,width=12.0cm,angle=270}
\vspace*{-1.2cm}
\caption{Measured spectra compared to computations with the interactions 
I,II,III (see text and table \ref{tab1}) relating to the $0^+$ ground
state of the core and IV relates to the $3^-$ core excitation. The
threshold of $E=0$ corresponds to infinite zero separation of the two
protons and the oxygen core.}
\label{fig0}
\end{figure}

Recently the decay products, two protons and $^{16}$O, from the
6.15~MeV $1^-$ state in $^{18}$Ne was measured \cite{gom01,ras08}. In
the same nucleus the 4.522~MeV $1^-$ state received attention as a
doorway state to produce the water molecule \cite{bel01} which has the
same number of neutrons, protons and electrons as the
$^{18}$Ne-atom. Also the first 3$^{+}$ state of this nucleus plays an
important role in astrophysics related to the abundance ratio between
$^{18}$F and $^{17}$F \cite{bar00}.  The $^{17}$F$(p,\gamma)^{18}$Ne cross section has been
investigated in a two-cluster coupled-channel model \cite{duf04}, and recently
also measured directly  \cite{chi09}.  The first question in this
connection is obviously which structures have these, and perhaps
other, $^{18}$Ne states. The low-energy spectrum of $^{18}$Ne is
typical for a quadrupole vibration with an equidistant spacing between
$0^+$, $2^+$, and a triplet of $0^+,2^+,4^+$ states \cite{til95}.
Still higher at and above the two-proton threshold a $0^+$ and a $3^+$
appear with a number of other states without an obvious recognizable
pattern, see Fig.\ref{fig0}.

The different states may originate from separate structures described
for example as vibrations, single-particle excitations, pairing
correlations, or obtained in combinations of even more complicated few
or many-body features. In particular the coupled two-body cluster model
provides one type of structure information \cite{duf04}. Also a number of interacting shell model
calculations have provided structure information about the low-lying
excited states of the $A=18$ isobaric system, $^{18}$O $^{18}$F
$^{18}$Ne, see e.g. \cite{she98,bro02}.  The results are in general
that many of the states are more complicated than two nucleons and the
$^{16}$O-core.  This is not surprising when the excitation energy is
sufficiently large to accomodate core excitations. However, the shell
model is designed to describe spatially confined bound state
structures without strong cluster configurations beyond the chosen
core-valence division.  This means that large-distance structures are
inaccessible or inaccurate in shell model calculations. This applies
in particular to resonances and doorway states in reactions.

Specifically, three-body decays of resonances cannot be described by two-body cluster or
shell models. The resonance structures necessarily change from the
many-body short-distance behavior to three-body clusters at
intermediate distances. In addition, the two-body structure is inadequate and the three-body structure itself
often change dramatically from intermediate to large distances
\cite{alv07,gar06,alv08,alv08a}.  The three-body structures must be 
accurately described to meet requirements of up-to-date measurements.
In other words shell model calculations necessarily must be
supplemented by few-body calculations as provided in the present paper.

To clear the road towards computing the three-body decays measured in
\cite{gom01,ras08} we start with assuming few-body structures to see
how far this will bring us in a quantitative description of the
various states.  Since $^{17}$Ne ($^{15}$O+p+p) is Borromean an extra
neutron suggest a four-body structure but a neutron and $^{15}$O form
a strongly bound doubly magic nucleus, $^{16}$O, and a three-body
structure of $^{16}$O+p+p is probably a better starting point.
Unfortunately it is then unlikely that the $1^-$ states simultaneously
are simple three-body structures maintaining an $^{16}$O ground state
core.  At least the $3^-$ state in the $^{16}$O core \cite{til95} can
be expected to contribute.

In the present paper we attempt to describe the low-lying (bound and
resonance) states in $^{18}$Ne as three-body states. If possible this
is a huge simplification from the full problem of 18 interacting
nucleons.  These investigations are a generalization of the classical
nucleon-core model and its extension to two mutually interacting
particles occupying single-particle levels provided by a core but
without additional nucleon-core interaction.  In any case, the
results are a prerequisite for description of three-body decaying
resonances like the measured $1^-$ state \cite{ras08,gom01}.  The
paper is organized as follows. In section \ref{sec2} we briefly
describe the notation by sketching the three-body method and the
constraints used to determine the crucial two-body interactions.  The
structures are shown in section \ref{sec4}, and the sizes and
electromagnetic transitions are given in \ref{sec5}. Finally, section
\ref{sec7} contains a summary and the conclusions.

\section{Method and interactions}
\label{sec2}

The principal model assumption is that $^{18}$Ne can be described as a
three-body system made by a $^{16}$O core and two protons. The wave
functions for the different bound states are obtained with the
hyperspherical adiabatic expansion method. A detailed description of
the method can be found in \cite{nie01}.

\subsection{Theoretical formulation}

This method solves the Faddeev equations in coordinate space. The wave
functions are computed as a sum of three Faddeev components
$\psi^{(i)}(\bm{x}_i,\bm{y}_i)$ ($i$=1,2,3), each of them expressed in
one of the three possible sets of Jacobi coordinates
$\{\bm{x}_i,\bm{y}_i\}$. Each component is then expanded in terms of a
complete set of angular functions $\{\phi_n^{(i)}\}$
\begin{equation}
\psi^{(i)}={1\over\rho^{5/2}} \sum_n f_n(\rho) \phi_n^{(i)}(\rho,\Omega_i);
(\Omega_i\equiv\{\alpha_i, \Omega_{x_i}, \Omega_{y_i} \}),
\label{eq1}
\end{equation}
where $\rho=\sqrt{x^2+y^2}$, $\alpha_i=\arctan({x_i/y_i})$,
$\Omega_{x_i}$, and $\Omega_{y_i}$ are the angles defining the
directions of $\bm{x}_i$ and $\bm{y}_i$.  Writing the Faddeev
equations in terms of these coordinates, they can be separated into
angular and radial parts:
\begin{eqnarray}
\hat{\Lambda}^2 \phi_n^{(i)}+\frac{2 m \rho^2}{\hbar^2} V_{jk}(x_i)
\left( \phi_n^{(i)} + \phi_n^{(j)}  + \phi_n^{(k)}   \right)  =
\lambda_n(\rho) \phi_n^{(i)} &&
\label{eq2} \\
\left[ -\frac{d^2}{d\rho^2} +  \frac{2m}{\hbar^2} (V_{3b}(\rho) - E)
+ \frac{1}{\rho^2}
\left( \lambda_n(\rho)+\frac{15}{4} \right) \right] f_n(\rho) \nonumber &&\\
& \hspace*{-8cm}
+ \sum_{n'} \left( -2 P_{n n'} \frac{d}{d\rho} - Q_{n n'} \right)f_{n'}(\rho)
= 0 &  \label{eq3}
\end{eqnarray}
where $V_{jk}$ is the two-body interaction between particles $j$ and
$k$, $\hat{\Lambda}^2$ is the hyperangular operator \cite{nie01} and
$m$ is the normalization mass. In Eq.(\ref{eq3}) $E$ is the three-body
energy, and the coupling functions $P_{n n'}$ and $Q_{n n'}$ are given
for instance in \cite{nie01}. The potential $V_{3b}(\rho)$ is used for
fine tuning to take into account all those effects that go beyond the
two-body interactions.
                                                                                
It is important to note that the set of angular functions used in the
expansion (\ref{eq1}) are precisely the eigenfunctions of the angular
part of the Faddeev equations. Each of them are in practice obtained
by expansion in terms of the hyperspherical harmonics. Obviously this
infinite expansion has to be cut off at some point, maintaining only
the most essential components. Specifically the contributing partial waves increase with energy and
distance. We include sufficiently many higher partial waves to reach
convergence.

The eigenvalues $\lambda_n(\rho)$ in Eq.(\ref{eq2}) enter in the
radial equations (\ref{eq3}) as a basic ingredient in the effective
radial potentials. Accurate calculation of the $\lambda$-eigenvalues
requires, for each particular component, a sufficiently large number
of hyperspherical harmonics. In other words, the maximum value of the
hypermomentum ($K_{max}$) for each component must be large enough to
assure convergence of the $\lambda$-functions in the region of
$\rho$-values where the $f_n(\rho)$ wave functions are not negligible.

Finally, the last convergence to take into account is the one
corresponding to the expansion in Eq.(\ref{eq1}). Typically, for bound
states, this expansion converges rather fast, and usually three or
four adiabatic terms are already sufficient.

\subsection{Proton-proton interactions}

The two-body low-energy scattering properties are crucial in the
description of weakly bound systems. The detailed short-distance
behavior is relatively unimportant.  Thus we adjust the parametrized
two-body interactions to known low-energy properties.  In the present
case this means the nucleon-nucleon interaction or, to be specific, the
proton-proton interaction.  We use a simple short-range potential
reproducing the experimental $s$- and $p$-wave proton-proton
scattering lengths and effective ranges. This assumes that effects of
the Coulomb interaction are removed from these scattering parameters.
Obviously the Coulomb potential is then afterwards added in the final
potential.  We assume the protons are point-like particles and the
Coulomb potential is then $e^2/r$.

The short-range nucleon-nucleon potential contains central, spin-orbit
($\bm{\ell}\cdot\bm{s}$), tensor ($S_{12}$) and spin-spin
($\bm{s}_1\cdot\bm{s}_2$) interactions, and is explicitly given as
\cite{gar04}
\begin{eqnarray}
\lefteqn{
V_{NN}(r) = 37.05 e^{(-(r/1.31)^2)}}  \nonumber \\  
 &&  -7.38e^{(-(r/1.84)^2)}
 -23.77e^{(-(r/1.45)^2)} \bm{\ell}\cdot\bm{s} \nonumber \\
  +& & \left(49.40e^{(-(r/1.31)^2)} +29.53e^{(-(r/1.84)^2)} \right) \bm{s}_1 \cdot \bm{s}_2 \nonumber \\
  & &    +7.16e^{(-(r/2.43)^2)} S_{12},
\end{eqnarray}
where $\bm{\ell}$ is the relative orbital angular momentum between the
two protons, and $\bm{s}=\bm{s}_1+\bm{s}_2$ is the total spin.  The
strengths are in MeV and the ranges in fm.  We shall refer to this
potential as the gaussian proton-proton potential. In actual
three-body computations we have tested, see e.g. \cite{rom08,gar03}, by using
other nucleon-nucleon potentials like the Argonne and the Gogny
potentials.  The three-body results were always indistinguishable.

\subsection{Proton-$^{16}$O  potential}
\label{sec p-16O pot}

The other two-body interaction is related to the proton-$^{16}$O
system.  The core, $^{16}$O, has intrinsic spin and parity $0^+$ and
the proton has spin $1/2$. The most general spin dependence is then of
spin-orbit form and each orbital angular momentum potential has the
form
\begin{equation}
V^{(\ell)}(r)=V^{(\ell)}_c(r)+V^{(\ell)}_{so}(r) \bm{\ell} \cdot \bm{s}_p
 + V_{C}(r)\;,
\label{eq5}
\end{equation}
where $\bm{\ell}$ is the proton-core relative orbital angular momentum
and $\bm{s}_p$ is the spin of the proton. These potentials should be
parametrized to reproduce low-energy scattering properties. We assume
gaussians for the radial shapes of all terms, i.e. central,
$V^{(\ell)}_c(r)$, and spin-orbit, $V^{(\ell)}_{so}(r)$.  The range
$b$ of the gaussians has to be related to the size of the
$^{16}$O-core.  We choose $ b = 2.60$~fm, as selected in \cite{gar03}
for the proton-$^{15}$O potentials.  The strengths are then left as
adjustable parameters.  The Coulomb potential, $V_{C}(r)$, can be
obtained either from $^{16}$O as a point particle or with a gaussian
charge distribution. For our purpose it suffices to use the potential
from a point charge as we did for the proton-proton Coulomb
interaction.

The most prominent features in low-energy scattering data are
reflected in properties of bound states and resonances, where the
dominant features in turn are energies of these structures.  We shall
therefore first aim at reproducing these energies. The two-body system
is $^{17}$F which has two proton bound states, i.e. a $1/2^+$-state at
$-0.105$ MeV and a $5/2^+$-state at $-0.600$ MeV measured relative to
the two-body threshold \cite{til95}.  The $s$-wave has no spin-orbit
term and the strength of $V^{(0)}_c(r)$ can be determined to reproduce
the energy $-0.105$ MeV. This should be the second $s$-state as the
first is occupied by core protons and consequently Pauli forbidden.
To exclude the lowest $s$-state in three-body computations we can
either use a shallow potential with only one bound $s$-state at
$-0.105$~MeV, or construct a phase equivalent potential (P.E.P) with
one less bound $s$-state \cite{gar99}.  For the $d$-state we keep the
same central potential strength we use for $s$-states while
adjusting the strength of the spin-orbit term to give a
$d_{5/2}$-state at $-0.600$~MeV.

\begin{figure}
\vspace*{-1.2cm}
\epsfig{file=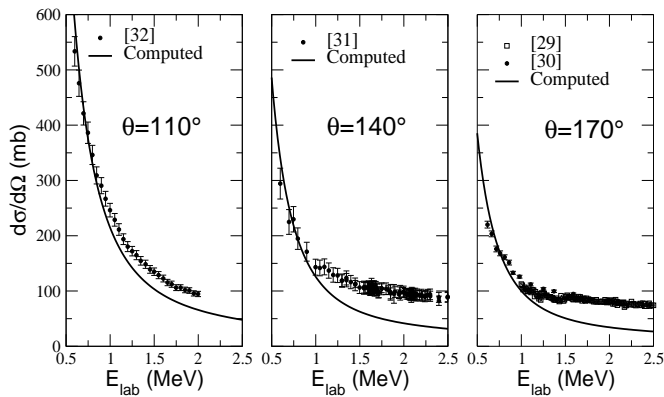,height=10cm,width=8cm,angle=270}
\vspace*{-1.6cm}
\caption{Differential Cross Sections for $^{16}$O$(p,p)^{16}$O. The computed lines were calculated using $s$ and $d$-waves generated with the potentials in table \ref{tab1}. }
\label{fig1z}
\end{figure}

These potentials now also lead to elastic cross sections, or
equivalently, phase shifts for each set of quantum numbers. We compare
in table \ref{tab1b} the computed phase shifts with measured values
from \cite{tra67}.  The $s_{1/2}$ and $d_{5/2}$ phase shifts are
matching the data perfectly as expected because the positions of the
bound states are well determined in our fits to match the measured
values. On the other hand the $d_{3/2}$ phase shifts deviate by
several degrees although both calculated and measured values are
small.  This partial wave is unimportant for the low-energy
structures, and we have not attempted any adjustment to these
observables.  We also computed the differential cross section as
measured for several angles in \cite{ami93,cho75,ram02,bra83}.  As we
can see in Fig. \ref{fig1z}, they are in perfect agreement with
results from calculations with our potentials including only $s$ and
$d$-waves for energies up to $1$~MeV where the higher partial waves
begin to contribute.  This is sufficient as the present computations
almost exclusively only need $s,p$ and $d$-waves. If more waves
occasinally are needed we use the same potential parameters as for the
$d$-waves.



To determine the $p$-wave two-body interaction the usual procedure
would be to reproduce negative parity $1/2$ or $3/2$-states. Such two
states are found above threshold in $^{17}$F at 2.504~MeV and 4.04~MeV
for $1/2$ or $3/2$, respectively, see \cite{liu90,mil01,fuk04}.
However, the sequence is opposite the established order from the
spin-orbit splitting.  Furthermore, these $p$-states should then
correspond to single-particle excitations into the $p-f$ shell which
first should appear at substantially higher energies.  Two choices
seem at first to be possible for the $p$-wave interaction. The first
is to enforce a $p$-wave potential to reproduce a $p_{1/2}$-energy at
2.504~MeV with the opposite sign of the spin-orbit potential perhaps
with a strength related to the $p_{3/2}$-energy at 4.04~MeV. The
second is to believe that these observed negative parity states are
complicated many-body states without any influence on the three-body
structure. This could be implemented by using the established shallow
$s$-potential with the spin-orbit term from the $d$-wave potential.

\begin{table}
\caption{Strengths in MeV of the central, 
$V_{c}^{(\ell)}(r)=S_c^{(\ell)}e^{-(r/b_c^{(\ell)})^2}$, spin-orbit,
$V_{so}^{(\ell)}(r)=S_{so}^{(\ell)}e^{-(r/b_{so}^{(\ell)})^2}$, and spin-spin,
$V_{ss}^{(\ell)}(r)=S_{ss}^{(\ell)}e^{-(r/b_{ss}^{(\ell)})^2}$,
potentials entering in Eq.(\ref{eq5}) for the four interactions used
in the calculations. The ranges are $b_c=b_{so}=b_{ss}=2.60$~fm in all
cases.  For the deep $s$-wave potentials the lowest $s$ state is
removed by construction of the phase equivalent potential (P.E.P.) 
from the central strength of the $d$-wave potential. The last column
refers to potentials built on the $3^-$ excited state of $^{16}$O. }
\begin{tabular}{|c|c|ccc|c|}
  \hline
          &                  &$ W_I$    & $W_{II}$  & $W_{III}$ & $W_{IV}$ \\  \hline 
$s$-waves & $S_c^{(\ell=0)}$ & P.E.P    & P.E.P     & $-27.26$ & $-28.50$   \\
       & $S_{ss}^{(\ell=0)}$ &          &           &          & $ 3.01$ \\  \hline
$p$-waves & $S_c^{(\ell=1)}$ & $-27.26$ & $-40.89$  & $-27.26$ & $-28.50$ \\   
       & $S_{so}^{(\ell=1)}$ & $-16.97$ & $5.89$    & $-16.97$ & $-10.00$ \\
       & $S_{ss}^{(\ell=1)}$ &          &           &          & $ 3.01$ \\  \hline
$d$-waves & $S_c^{(\ell=2)}$ & $-87.27$ & $-87.27$  & $-87.27$ & $-73.07$  \\  
       & $S_{so}^{(\ell=2)}$ & $-16.97$ & $-16.97$  & $-16.97$ & $-10.00$ \\ 
       & $S_{ss}^{(\ell=2)}$ &          &           &          & $ 3.01$ \\  \hline
\end{tabular}
\label{tab1}
\end{table}

\begin{table}
\caption{Phase Shifts for $s$ and $d$-waves of $^{17}$F. For each energy the first row shows the experimental values from \cite{tra67} and the second row gives the computed phase shifts with potential I.}
\begin{tabular}{|c|c|c|c|}
  \hline
 E$_{c.m.}$& $\delta_{s_{1/2}}$&  $\delta_{d_{3/2}}$&    $\delta_{d_{5/2}}$ \\ \hline
2.32 & 145.0 & 2.4 & 179.2   \\
     & 142.7 & 0.6 & 179.0 \\
2.42 & 142.9 & 2.9 & 178.9 \\  
     & 140.7 & 0.7 & 178.9 \\
2.55 & 140.1 & 3.6 & 178.6 \\
     & 138.2 & 0.8 & 178.8 \\
2.80 & 135.2 & 5.0 & 178.0 \\ 
     & 133.6 & 1.2 & 178.5 \\ \hline
\end{tabular}
\label{tab1b}
\end{table}

There is also a third option, which relates these levels to the $3^-$
excited state of the $^{16}$O-core at 6.13~MeV above the ground state.
This core-state coupled to a $5/2^+$ proton single-particle state
could produce the two observed $1/2^-$ and $3/2^-$ states. A similar
coupling to a $1/2^+$ proton state produce $5/2^-$ and $7/2^-$ states.
The core excited $1^-$ state at $7.12$~MeV could also couple to the
$5/2^+$ and $1/2^+$ single-particle states to give the $3/2^-$,
$5/2^-$, $7/2^-$ and the $1/2^-$, $3/2^-$ states.  Then the
lowest-lying $1/2^-$ state would not be related to the lowest $d$-wave
but to the higher lying $s$-wave.  The $3/2^-$ would be a mixture of
both $s$ and $d$-waves.  We restrict ourselves to explore the simplest
combination relating to the $3^-$ excited core state.

These negative parity states could then have contributions from both
the core excited $3^-$ state and the $0^+$ ground state of the
$^{16}$O core.  To the degree that they are decoupled in the two-body
states they would also be decoupled in the three-body states.
Furthermore, the lowest possible partial waves for most of the
low-energy three-body $J^{\pi}$ states correspond uniquely to either
$3^-$ or $0^+$ core states. Other contributions are less favored by
either relative energy or core excitation. Thus decoupling on the
three-body level could be rather well fulfilled.

We have now established several options for the two-body proton-core
potential. The strengths of the resulting different potentials are
given in table~\ref{tab1}. We include specific interactions for $s$,
$p$, and $d$-waves. In all cases are the two bound state energies
reproduced.  Potentials I and II employ the same deep potential for
both $s$ and $d$-waves, the phase equivalence is used for $s$-waves.
Potential III maintain the $d$-wave from I and II, whereas a shallow
potential with one bound state is used for $s$-waves. The $p$-wave
potentials I and III use the central $s$-potential from III and the
spin-orbit from I and II. In II the $p$-wave is adjusted to give the
measured $p_{1/2}$ energy at $2.504$~MeV.  The most likely candidate
as spin-orbit partner of the 1/2$^-$ state is a $3/2^-$ resonance at
4.04 MeV (above threshold). Therefore we adjust the central and
spin-orbit depth to fit these energies while keeping the same
configuration as I for $s$ and $d$-waves.

Potentials I, II, and III are based on the spin zero ground state of
$^{16}$O. Potential IV is based on the $3^-$ excited state of $^{16}$O
where the $d$-wave is the decisive component in the description of the
low-lying states. This finite spin of the core requires the
generalization of the potential to include the spin-spin term, i.e.
\begin{equation}
V^{(\ell)}(r)=V^{(\ell)}_c(r)+V^{(\ell)}_{so}(r) \bm{\ell} \cdot \bm{s}_p+V^{(\ell)}_{ss}(r) \bm{s}_c \cdot \bm{j}_p +V_C(r),
\label{eq5b}
\end{equation}
where we maintain the gaussian shapes of range $2.60$~fm of the radial
potentials, and the Coulomb potential again is for point particles as
for potentials I, II and III.  We use the energies of the $1/2^-$ and
$3/2^-$ states to determine the central and spin-spin strengths for
the $d_{5/2}$-wave coupled to $3^-$ of the core. This leaves the
spin-orbit strength as a free parameter which is chosen to be similar
in size to the values of the other potentials. It only has to be large
enough to place the $d_{3/2}$-waves at sufficiently high energies to
make their effects negligibly small.

Coupling of $s_{1/2}$-waves and $3^-$ produce $5/2^-$ and $7/2^-$
states which however also arise from the $d_{5/2}$ couplings. These
relatively high-lying states are expected to contribute very little to
the low-lying three-body states.  We use the energy, $3.257$~MeV above
threshold, of the $5/2^-$ resonances to estimate the central strength
of the $s$-wave potential while maintaining the spin-spin strength
derived from the $d$-waves.  Any $p$-wave would in principle mix into
the positive parity states but both angular momentum and energy
indicate negligibly small effect. We have kept the same potential as
for $s$-waves adding the spin-orbit from $d$-waves.

We have not attempted to reproduce the widths of these resonances as
that would require more parameters like variation of the range of the potentials.

\section{Structure}
\label{sec4}

The angular eigenvalues obtained from eq.(\ref{eq2}) enter in the
coupled set of radial equations eq.(\ref{eq3}) as the crucial ingredient
of the effective potentials $V^{(n)}_{eff}(\rho)=\frac{1}{\rho^2}
\left(\lambda_n(\rho)+\frac{15}{4} \right)$.  To solve the angular part  
of the Faddeev equations we use expansion of each Faddeev component on
hyperharmonic wavefunctions.  The angular wavefunctions are expressed
in the Jacobi coordinates, $\bm{x}$ and $\bm{y}$, where the
corresponding orbital angular momentum quantum numbers, $\ell_x$ and
$\ell_y$, couple to the total orbital angular momentum $L$.  The
parity $\pi$ is then given by the odd or even character of $\ell_x +
\ell_y$.  The spins of the two particles connected by the $\bm{x}$
coordinate couple to $s_x$, that in turn couples with the spin of the
third particle to the total spin $S$.  Finally $L$ and $S$ couple to
the total angular momentum $J$ of the three-body system.  The last
quantum number of the basis is the hypermomentum $K=2n+\ell_x+\ell_y$
where the non-negative integer $n$ is the number of nodes.  For each
set of angular quantum numbers we include all $K$-values smaller than
a given $K_{max}$ chosen to guarantee convergence for all necessary
hyperradii.

\begin{figure}
\epsfig{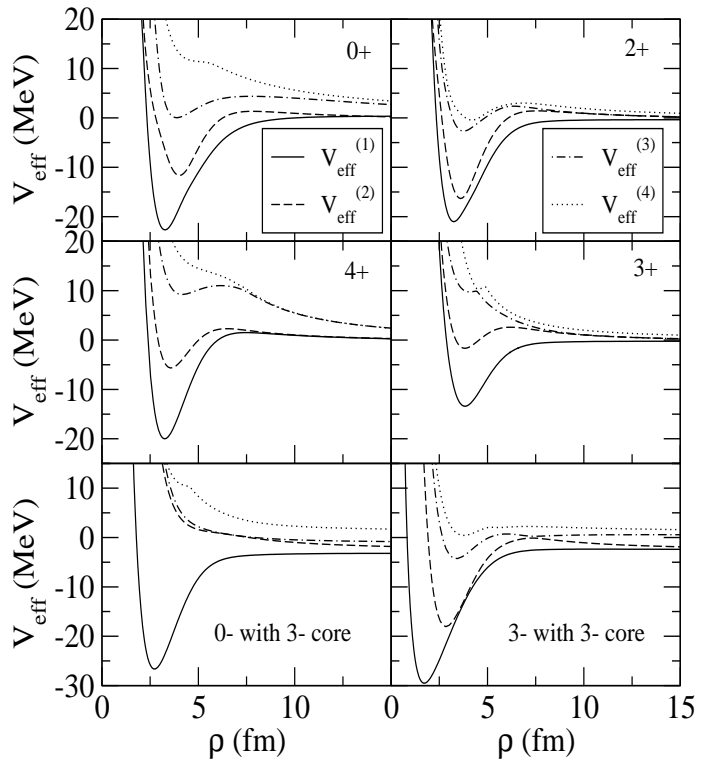}
\caption{The four lowest adiabatic effective potentials $V^{(n)}_{eff}$ for 
positive parity states in $^{18}$Ne built on the $0^{+}$ core
structure and with total angular momentum $J^{\pi} = 0^{+}, 2^{+},
3^{+}, 4^{+}$ and also $J^{\pi} = 0^{-},3^{-}$ for $3^{-}$ core. The proton-core potential is I for the $0^{+}$ core cases and potential IV for the $3^{-}$ core case as given in
table~\ref{tab1}.}
\label{fig1a}
\end{figure}

\subsection{Effective potentials}

The potentials in eq.(\ref{eq3}) determine the structure of the
$^{18}$Ne states. We show in Fig.~\ref{fig1a} the lowest effective
potentials for the bound states and resonances selected among
the different structures we have investigated, i.e. $J^{\pi} =
0^{\pm}, 1^{\pm},2^{\pm}, 3^{\pm}, 4^{+}$ built on both 0$^{+}$ and
3$^{-}$ core-structures. We see attractive pockets in the two lowest
potentials for the positive parity states built on the 0$^{+}$
core. These potentials are strongly influenced by the bound $s$ and
$d$-states in the proton-core potentials. At larger distances they
asymptotically approach the energies, $-0.6$~Mev or $-0.1$~Mev, of the
$^{17}$F bound states.  This reflects the large-distance structure of
one proton far away from the $^{17}$F nucleus in the corresponding
bound states, i.e. ground state, $d_{5/2}$, and first excited state,
$s_{1/2}$.

For the total angular momentum, $0^{+}$, the last proton is in either
$d$ or $s$-waves around the corresponding $^{17}$F structures. The
$2^{+}$ potentials can have $^{17}$F nucleus in the $d_{5/2}$-state
surrounded by the second proton with an angular momentum of either
$0,2$ or $4$ in total allowing $5$ different couplings.  We only
include the three lowest of these potentials which approach
$-0.6$~MeV.  The fourth of the $2^{+}$ potentials in Fig.~\ref{fig1a}
approaches $-0.1$~MeV.  It corresponds to the $^{17}$F nucleus in the
$s_{1/2}$-state where the second proton has angular momentum $2$
around $^{17}$F subsystem.  At small distances two of these potentials
exhibit rather appreciable attraction.

The $4^{+}$ potentials allow proton angular momenta of $2,4,6$, and
$4$ around the $^{17}$F nucleus in $d$ or $s$-states, respectively.
We find that two of these potentials approach $-0.6$~MeV and one
approaches $-0.1$~MeV at large distance.  For the $3^{+}$ potentials
the angualr momentum combinations are $2,4,6$, or $2,4$, for the bound
$d$ and $s$-states, respectively. The lowest potentials again approach
$-0.6$~MeV and $-0.1$~MeV.  Most of the higher-lying potentials
approach zero reflecting a genuine three-body continuum structure.


The lowest effective potentials for negative parity states built on the
$0^{+}$ core structure can also be computed with the help of the
single-particle $p$-waves.  Using $p$-wave interaction from the $s$ and
$d$-waves in potential I we find again relatively attractive pockets in
the lowest adiabatic potentials for $1^{-}$ and $3^{-}$ but they are almost totally
absent for $0^{-}$ and $2^{-}$. The large-distance approaches are found in
all cases towards the $s$ and $d$-wave two-body structures.  Bound states
or resonances of corresponding structures may then arise for $1^{-}$ and
$3^{-}$.



The potentials for the $0^-$ and $3^-$ states built on an excited
$3^-$ $^{16}$O-core state are also shown in Fig.\ref{fig1a}.  Similar
but less attractive potentials are found for 1$^-$ and $2^-$. The
threshold energy for all these states is then with respect to this
core excited state at $6.13$~MeV above the $0^+$ ground state of
$^{16}$O.  The two-body states used to adjust the interactions are
bound with respect to this core excited state.  These three-body
potentials also exhibit attractive pockets at small distances. Their
large-distance asymptotics also reflect these two-body ``bound
states'' where the three lowest potentials approach $-3.6$~MeV,
$-2.9$~MeV and $-2.1$~MeV.  These values correspond to the proton-core
``bound'' states of $(\ell,S)=(2,5/2),(0,5/2),(2,7/2)$ where $\ell$ is
the orbital angular momentum and $S$ the total spin quantum number
including the $3^-$ from the core.


At small distances several attractive pockets appear. For $0^-$ only
one deep and broad potential can bind with respect to the $3^-$
excited core state.  This is essentially due to an even combination of
the $(\ell,S)=(2,5/2),(0,5/2)$ partial waves.  For $1^-$, $2^-$ and
$3^-$ two attractive potentials are found with varying relative
depths.  They are mostly constructed from those combinations of
$(\ell,S)=(2,5/2),(0,5/2),(2,7/2)$ which allow spatially overlapping
antisymmetric two-proton states.

\subsection{Three-body energies}

For each set of adiabatic potentials we solve the coupled set of
radial equations in eq.(\ref{eq3}).  The resulting eigenvalues are
shown in tables~\ref{tab6a} and \ref{tab6b} for bound and unbound
solutions, respectively.  If the energies are decisive for
applications, as for breakup and decaying resonances, we can fine tune
by use of the effective three-body potentials, $V_{3b}(\rho)$, in
eq.(\ref{eq3}). This would maintain the structure essentially
completely unchanged.  Such adjustments are not included in the
eigenvalue tables.

\begin{table}
\caption{Spectrum of the positive parity bound states in $^{18}$Ne
for the different proton-core interactions in table~\ref{tab1}. The
numerical results have been obtained without inclusion of a three-body
potential in Eq.(\ref{eq3}).  The two-proton separation energies are
given in MeV.  The last column gives the available experimental
energies from \cite{til95}.}
\begin{tabular}{|c|cccc|}
\hline
        & $W_I$ & $W_{II}$ & $W_{III}$ &   $E_{exp}$   \\ \hline
$0^+_1$ & $-4.30$   &  $-5.06$   &   $-5.32$       &$-4.52$  \\
$2^+_1$ & $-2.81$   &  $-3.19$   &   $-3.32$      &$-2.63$  \\
$4^+$   & $-1.24$   &  $-1.17$   &   $-1.14$      &$-1.14$  \\
$0^+_2$ & $-0.46 $  &  $-0.49$   &   $-1.38$       &$-0.94$  \\
$2^+_2$ & $-0.89 $  &  $-0.91$   &   $-0.88$       & $-0.90$  \\
$3^+$   & $-0.08 $  &  $-0.11$   &  $ -0.02$       & $+0.04$  \\ \hline
\end{tabular} 
\label{tab6a}
\end{table}

The unbound states are decaying resonances with a width arising from
cluster, or equivalently two-proton, emission. To compute such
continuum states complex scaling could be applied.  States built on
the excited $3^-$ core state with energies less than 6.13~MeV are
bound states in the computation and complex scaling is not needed.
They can only decay electromagnetically, or by the neglected coupling
to the $0^+$ ground state.  In any case we focus in this paper only on
the real part of the energies which above their respective thresholds
are computed in two steps. First a sufficiently attractive three-body
potential is added to bind the state. Second the strength of that
potential is varied and the resulting energy is extrapolated to the
estimated energy obtained for zero strength. This is the so called 
analytic continuation of the coupling constant method \cite{tan97}, 
which is especially
relevant for states arising from the potentials for negative parity states based on the $0^+$ core.

The bound states shown in table~\ref{tab6a} for different potentials
are quite stable independent of choice of potential.  In all cases the
level ordering is reproduced and the energies are also rather close to
the measured values. We only adjusted the potentials to two-body bound
state and resonance properties with the simplest possible radial
shapes.  Potential I leads in general to less binding but closer to
measurements than the two other potentials. The deviations are less
than $200$~keV except for the $500$~keV underbinding of the last $0^+$
state.  

In potential I the phase equivalent $s$-wave potential produces a
repulsive core at short distance and the valence protons are pushed
away from the center. This implies that potential III should bind more
when proton-core $s$-wave configurations are substantial as for the
first $2^+$, the $3^+$, and both the $0^+$ states.  The energies of
both the second $2^+$ and the $4^+$ states do not depend on the chosen
potential but matches almost perfectly the measured values.  The $3^+$
state is weakly bound by less than about $100$~keV in the computation
in agreement with the measured value close to the breakup threshold.

The results from potential I and II deviate surprisingly much for the
two lowest states indicating that the $p$-wave components play a role.
Then the description from potential II implies that these states have
contributions of proton-core single-particle $p_{1/2}$ character.
This is inconsistent with the shell structure of the core-nucleus. We
prefer potentials I and III with the much weaker $p$-wave attraction.
Potential I fits the experimental values better than potential II.

As shown in table~\ref{tab6b} we find a $3^-_2$ resonance at about $3.10$~MeV and a low-lying
$1^-_1$ resonance at about $0.2-0.4$~MeV.  Both states are based on
the $0^+$ core ground state.  These values are more uncertain since
they are obtained by extrapolation with a strongly attractive
three-body potential.  These structures are not present for $0^-$ and
$2^-$ as already seen from the disappearance of attractive pockets in
the lowest adiabatic potentials for negative parity states based on the $0^+$ core.

An alternative to potential II in descriptions of the negative parity
states is potential IV.  We show several of the resulting energies in
table~\ref{tab6b}.  In particular there appears a $1^-_2$ state built
on the excited $3^-$ core state and bound compared to this state by
$3.98$~MeV implying that it is an observable resonance at about
$2.12$~MeV.  Since the energies of the two $1^-$ states differ by
$2$~MeV they may be decoupled in practice. The $0^-$ energy is about
$-5.45$~MeV with respect to the core excited state and therefore at an
energy of $0.58$~MeV above the two-proton threshold.

We find another $3^-$ ``bound" state at about $-3.09$~MeV
corresponding to $1.33$~MeV.  For $2^-$ we find two ``bound states at
about $-2.92$~MeV and $-2.00$~MeV corresponding to $3.41$~MeV and
$4.13$~MeV.  The second of these is a resonance in the two-body
continuum of the bound proton-$^{16}$O($3^-$) system and the other
proton. There is no confining barrier and it easily leaks out
corresponding to a large width.  All these negative parity states can
easily be matched to measured energies in the continuum. However, such
a comparison is not very revealing due to the inevitable inaccuracy
from the three-body approximations.  At least more structure
information is needed.  Still we show some of the lowest measured
values in table~\ref{tab6b}.  Many higher-lying levels are found
experimentally.

\begin{table}
\caption{Spectrum of the unbound, mostly negative parity, states
in $^{18}$Ne for different proton-core interactions in
table~\ref{tab1}.  The numerical results have been obtained without
inclusion of a three-body potential in Eq.(\ref{eq3}).  The two-proton
separation energies are given in MeV with respect to the threshold for
separation into two protons and the core in its ground state. These
energies are given below the potentials used in the computation. The
fifth and sixth columns give the available experimental energies and
spin and parities, if known, from \cite{til95}.  The computations with
potential IV are for a core-excited state at 6.13~MeV implying that
these states with energies lower than 6.13 MeV behave as bound states. }
\begin{tabular}{|c|ccc|cc|}
\hline
$J^{\pi}$ & $W_{I}$ & $W_{III}$ & $W_{IV}$ & $E_{exp}$ & $J^{\pi}_{exp}$
\\ \hline
$1^-_1$ & $0.44$   & $0.20$    &                & $0.00$      & $1^-$ \\
$0^-$    &              &               &  $0.58$     & $0.57$     &
$(2^+,3^-)$ \\
  &            &                &       &  $0.63$   &  $(2^+,3^-)$  \\
$3^-_1$  &            &                & $1.33$      &  $0.93$   &   \\
$1^-_2 $ &            &                & $2.12$      &  $1.63$   &  $1^-$ \\
$3^-_2$  & 3.10     & 3.10         &                &     &    \\
$2^-_1$  &            &                 & $3.41$     &    &   \\
$2^-_2$  &            &                 & $4.13$     &    & \\  \hline
\end{tabular}
\label{tab6b}
\end{table}

\subsection{Wavefunctions}
\label{sec3}

The eigenfunctions are found as expansion coefficients on the
hyperharmonic basis with quantum numbers for each of the Faddeev
components, i.e. ($\ell_x,\ell_y,L,s_x,S,J$).  Each eigenfunction can
be expressed in one set of Jacobi coordinates with corresponding
probabilities depending on potential and quantum numbers. We shall in
this section in details discuss the two bound 0$^{+}$ states and the
1$^{-}$ resonance located close to the threshold.  For completeness we
give the decompositions of the other states in the appendix.

\begin{table}
\caption{Components included in the calculations for the 0$^+$ states. 
The upper part correspond to the first Jacobi set ($\bm{x}$ between
the two protons). The lower part corresponds to the second and third
Jacobi sets ($\bm{x}$ from core to proton). The 6$^{th}$ column gives
the maximum value of the hypermomentum used for each component. The
last three columns give the contribution from each component to the
0$^+$ wave functions for potentials I, II and III respectively. Only contributions larger than 0.1 is given. The two
numbers for each component correspond to the $0^+_1$ and $0^+_2$
states, respectively. }
\begin{tabular}{|cccccc|ccc|}
\hline
$\ell_x$ & $\ell_y$ & $L$ & $s_x$ & $S$ & $K_{max}$ &$W_I$  & $W_{II}$ & $W_{III}$  \\
\hline
   0  &  0  &  0  &  0  &  0 & 120 &  80.7  &  81.0  &  78.9   \\
      &     &     &     &    &     &  82.0  &  82.2  &  86.6    \\
   1  &  1  &  1  &  1  &  1 &  90 &  17.0  &  17.0  &  17.1    \\
      &     &     &     &    &     &   8.9  &   8.5  &   9.7    \\
   2  &  2  &  0  &  0  &  0 &  90 &   2.3  &   2.1  &   3.9    \\
      &     &     &     &    &     &   9.1  &   9.3  &   3.7   \\ \hline
   0  &  0  &  0  & 1/2 &  0 & 120 &  24.9  &  21.9  &  44.4   \\
      &     &     &     &    &     &  74.3  &  76.0  &  53.1    \\ 
   1  &  1  &  0  & 1/2 &  0 &  90 &   5.0  &  13.2  &   0.4    \\
      &     &     &     &    &     &   0.4  &   0.3  &   4.3    \\ 
   1  &  1  &  1  & 1/2 &  1 &  85 &   0.2  &   1.5  &   0.2    \\
      &     &     &     &    &     &   0.2  &   0.3  &   0.1    \\ 
   2  &  2  &  0  & 1/2 &  0 & 100 &  52.6  &  47.5  &  37.4    \\
      &     &     &     &    &     &  16.1  &  14.9  &  32.6    \\ 
   2  &  2  &  1  & 1/2 &  1 &  90 &  17.3  &  15.9  &  17.6   \\
      &     &     &     &    &     &   9.1  &   8.5  &   9.8  \\ \hline
\end{tabular}
\label{tab3}
\end{table}

\subsubsection{Bound states}

The available single-particle states for the two protons are $d_{5/2}$
and $s_{1/2}$ orbits. Two-particle states of both protons in $d_{5/2}$
produce the sequence of $0^+$, $2^+$, and $4^+$ states where the odd
angular momenta are forbidden due to the antisymmetry
requirement. Using two $s_{1/2}$ states we can only produce a $0^+$
state. One proton in each of the $d_{5/2}$ and $s_{1/2}$ states
produce one $2^+$ and one $3^+$ state.  As seen in Fig.~\ref{fig0}
they all appear in the computed spectrum.  The partial wave
decomposition reveal the microscopic structure of the states.

 We show the partial wave decomposition
in table~\ref{tab3} for the two $0^+$ states.  The upper part using
the first Jacobi system, where the $\bm{x}$-coordinate connects the
two protons, exhibits only small variation between results from the
different potentials, the $s$-waves of about $80$\% dominate for both
states in the three cases.

The picture is very different in the proton-core Jacobi system where
the variation with potential is larger. 
The $0^+_2$ state with potential I has about $74\%$ in the $s_{1/2}^2$
configuration which essentially means that both protons are in $s_{1/2}$-states. 
With potential III the components $d_{5/2}$ and $s_{1/2}$ are more even, i.e. about $55\%$
$d_{5/2}^2$ in $0^+_1$ and $53\%$ $s_{1/2}^2$ in $0^+_2$.  This
reflects the lack of repulsion at short distance in the $s$-wave
interaction which favor $s$-waves in the ground state.
With potential II we still get about $76\%$ $s_{1/2}^2$ configuration
in the $0^+_2$ state. We also included components with $\ell_x$ and $\ell_y$ larger than 2 although their contributions are found to be negligible after the computations as we already mentioned in section \ref{sec p-16O pot}.

The configurations obtained here for the $0^+_2$ state essentially
only contains $sd$-waves.  Early shell model calculations of both
mirror nucleus $^{18}$O \cite{law74} and $^{18}$Ne \cite{she98} gave
about $33\%$ and collective motion the remaining $67\%$.  Our computed
energy only deviates from measurements by about $0.4$~MeV, see
table~\ref{tab6a}, which is a typical deviation in such three-body
calculations.  This is therefore surprising if $2/3$ of the structure 
should have a completely different origin. It is more likely that the
$sd$-configurations contribute by more than $1/3$ to the $0^+_2$
structure. This may be reconciled with the shell model results if part
of the collective motion also is of $sd$-character.

The other four bound states of $J^{\pi}=2^+,3^+,4^+$ are also
decomposed in partial wave configurations and shown as tables in the
appendix. Both $2^+$ and $3^+$ states consist of proton-core $s$ and
$d$-waves, and the $4^+$ state of solely $d$-waves.

\subsubsection{Unbound states}
\label{sec3c2}

With potential II it is a priori not excluded to find negative parity
energies with resemblance to the measured spectrum.  The corresponding
structures are on the other hand not expected to reproduce measured
properties.  The basic problem is the assumption of a single-particle
$p_{1/2}$-state in the low-lying spectrum of $^{17}$F. Potential IV,
built on the core excited $3^-$ state is in general expected to
provide better structure properties. Then a number of $0^-,1^-,2^-,$
and $3^-$ states should arise as combinations of $d$ and $s$ states
coupled to the $3^-$ core state.  However, it is here worth
emphasizing that there might be different, perhaps essentially
uncoupled, structures of the same $J^{\pi}$ but built on different
core states. We find such states with $J^{\pi}=1^-$ and $3^-$.

\begin{table}
\caption{The same as table \ref{tab3} for the 1$^-_1$ state in $^{18}$Ne
with the 0$^{+}$ core. }
\begin{tabular}{|cccccc|cc|}
\hline
$\ell_x$ & $\ell_y$ & $L$ & $s_x$ & $S$ & $K_{max}$ & $W_{I}$ &
$W_{III}$\\ \hline
 0  &  1  &  1  &  0  &  0  &  120 &  82.9  &80.4    \\
 1  &  0  &  1  &  1  &  1  &   80 &   6.0  &8.1    \\
 2  &  1  &  1  &  0  &  0  &   80 &   5.3  &5.7    \\
 1  &  2  &  1  &  1  &  1  &   80 &   3.5  &2.7    \\
 1  &  2  &  2  &  1  &  1  &   80 &   2.3  &3.1    \\  \hline
 0  &  1  &  1  &  1/2  &  0  &  100 & 19.4 &19.6      \\
 0  &  1  &  1  &  1/2  &  1  &   80 &  1.4 &1.6     \\
 1  &  0  &  1  &  1/2  &  0  &  100 & 19.1 &21.2      \\
 1  &  0  &  1  &  1/2  &  1  &   80 &  0.9 &1.1     \\
 2  &  1  &  1  &  1/2  &  0  &  100 & 29.7 &22.4      \\
 2  &  1  &  1  &  1/2  &  1  &   80 &  5.0 &4.7     \\
 2  &  1  &  2  &  1/2  &  1  &   80 & 1.9  &2.1      \\
 1  &  2  &  1  &  1/2  &  0  &  100 & 21.8 &21.0      \\
 1  &  2  &  1  &  1/2  &  1  &   80 &  4.3 &4.4     \\
 1  &  2  &  2  &  1/2  &  1  &   80 &  1.7 &1.9      \\    \hline
\end{tabular}
\label{tab9a}
\end{table}

The first $1^-_1$ state is built on the $0^+$ core ground state. The
cluster model $^{14}$O+$\alpha$ in \cite{duf04} cannot describe this
state which is suggested as a candidate for burning $^{18}$Ne into
water while releasing a lot of energy \cite{bel01}.  The decomposition
shown in table~\ref{tab9a} is in the first Jacobi system seen to be
dominated by $s$-wave components between the two protons. In the other
Jacobi system both proton-core $s$ and $d$-waves contribute about
$20\%$ and $p$-waves by twice that amount. In this way the attraction
of the two interactions, $s_{1/2}$ and $d_{5/2}$, are optimized.  The
lowest adiabatic potential contributes by $93\%$.

\begin{table}
\caption{The same as table \ref{tab3} for the second 1$^-_2$ state in 
$^{18}$Ne with the 3$^{-}$ core excited state.  }
\begin{tabular}{|cccccc|c|}
\hline
$\ell_x$ & $\ell_y$ & $L$ & $s_x$ & $S$ & $K_{max}$ & $W_{IV}$  \\ \hline
   2  &  0  &  2  &  0  &  3  &   60 &     1.7    \\
   0  &  2  &  2  &  0  &  3  &   60 &     3.3    \\
   1  &  1  &  2  &  1  &  2  &  100 &    89.4    \\
   1  &  1  &  2  &  1  &  3  &   60 &     4.3    \\
   2  &  2  &  2  &  0  &  3  &   60 &     1.3    \\  \hline
   0  &  2  &  2  &  5/2  &  2  &  100 &   44.6      \\
   0  &  2  &  2  &  5/2  &  3  &  100 &    0.2     \\
   2  &  0  &  2  &  5/2  &  2  &  100 &   45.2     \\
   2  &  0  &  2  &  5/2  &  3  &  100 &    9.4      \\
   1  &  1  &  2  &  5/2  &  2  &  100 &    0.6     \\   \hline
\end{tabular}
\label{tab9}
\end{table}


The second $1^-_2$ state is built on the $3^-$ excited core state.
The three-body decay of this state into two protons and the ground
state of $^{16}$O is measured and analysed in terms of sequential,
virtual sequential and direct decay branching ratios
\cite{ras08,gom01}.  Such decay must take place through couplings to
other states as this state is bound with respect to the $3^-$ core
excited state.  The decomposition in table~\ref{tab9} show dominance
of $p$-waves in the first Jacobi system. In the other Jacobi systems
this results in equal amounts of $s$ and $d$-waves in the proton-core
subsystem. This implies that the proton-core spin has to be $5/2$.
The lowest and second potential contribute by $88\%$ and $11\%$,
respectively.

We find several differences with respect to the shell model results
in~\cite{bro02} where the structure of the different $1^-$ states in
$^{18}$Ne are discussed in detail.  The first five of these shell
model $1^-$ states have for one choice of interactions respectively
about $2\%$, $20\%$, $6\%$, $12\%$, $49\%$ of configurations with
$^{16}$O in the ground state.  For comparison our two $1^-$ states
contain $100\%$ either ground or $3^-$ excited state of $^{16}$O. The
shell model results strongly indicate mixing of different excited
states of $^{16}$O coupled to the two protons.  However, these shell
model structures at most determine the short-distance behavior whereas
the intermediate and large-distance structures can be completely
different and hence also the resulting momentum distributions after
decay.

The partial wave decomposition of all other computed unbound states
are shown in the appendix. The $0^-$ and the two $2^-$ built on the
core excited $3^-$ state all consist of proton-core $s$ and $d$-waves.
The two $3^-$ states consist of proton-core $p$-waves and $p$ and
$d$-waves when built on the $3^-$ and $0^+$ core states, respectively.

The partial wave decomposition focus on the angular structure.  The
structures can be further characterized by the overlaps between
valence wavefunctions of negative and positive parity states.  Here it
is necessary to remember that the core structure differs, and true
overlaps are zero. However, the valence part may have contributions of
precisely the same partial waves which in turn has to be coupled to
$0^+$ or $3^-$ to give the different total angular momenta.  The
overlaps can be estimated from the partial wave decompositions in the
tables. These angular overlaps should be multiplied by radial overlap
functions which in general are rather similar for low-lying
states. The orbital and spin angular momentum couplings introduce in
some cases another substantial reduction factor.  It is in this way
easily seen that the two $0^+$ states and the $3^-$ state have
the largest overlaps both exceeding $0.6$. Also the $3^+$ and the
$0^-$ states seem to overlap by more than $0.4$ similar to the first
$2^+$ and the first $2^-$ states.  All other overlaps are rather
small.

\section{Moments and transition probabilities}
\label{sec5}

The expectation values of the operators provide observables for each
state. The most interesting are those related to sizes and
lifetimes.  We give details in the next two subsections.

\subsection{Relative sizes}

The sizes are observable quantities, where the simplest are the second
moments of charge and matter distributions.  These are the root mean
square radii which often only are available for the ground state. The
charge radius is most accurately obtained by electromagnetic probes
like electrons. The matter radius is for light nuclei derived from
measurements of interaction cross sections. The excited states are
closer to the threshold for breakup and therefore more likely to
develop a spatially extended halo structure.  This would have
observable implications for breakup cross sections.  Effects of
binding energy and angular momentum are both important \cite{jen04}.
The three-body results are related to observables by including the
finite extension of core distribution for matter and charge.  In the
present case we have for the matter distribution
\begin{eqnarray} \label{e77} 
 \langle r^2 \rangle &=&  \frac{16}{18} r^2_{core} + 
 \frac{1}{18}\langle \rho^2 \rangle \; , \\  \nonumber 
 \langle \rho^2 \rangle &=& \frac{1}{2}\langle r^2_{pp} \rangle +
 \frac{32}{18} \langle r^2_{c,pp} \rangle 
 = \frac{16}{17}\langle r^2_{pc} \rangle +
 \frac{17}{18} \langle r^2_{p,cp} \rangle \;,
\end{eqnarray}
where $r^2_{core} $ is the mean-square radius of the core.  For the
charge distribution we get
\begin{eqnarray} \label{e84} 
 \langle r^2 \rangle_{ch} &=&  \frac{8}{10} r^2_{core,ch} + 
 \frac{1}{20}\langle r^2_{pp} \rangle +
 \frac{136}{810} \langle r^2_{c,pp} \rangle   \; ,  \\ \label{e87} 
  \frac{ \langle r^2 \rangle_{ch}}{\langle r^2 \rangle} &=& 
 \frac{9}{10}  \frac{r^2_{core,ch} + \langle r^2_{pp} \rangle/16 +
 17 \langle r^2_{c,pp}\rangle_{ch}/81}
 {r^2_{core} + \langle r^2_{pp} \rangle/32 + 
 \langle r^2_{c,pp}\rangle_{ch}/9} \;,\;
\end{eqnarray}
where $r^2_{core,ch} $ is the mean-square radius of the core charge
distributions. We assume that charge and matter distributions are
identical for the core. With these expressions we can always insert a
different value for the core moments if better parameters become
available.

\subsubsection{Bound states}

The root-mean-square, charge and matter, radii for the computed bound
states are given in table~\ref{tab11}.  The results are essentially
independent of the core-proton potential but the trends reflect that
smaller binding energies give larger radii and viceversa.  This is
especially clearly seen for the $0^+_2$ state where potential III
gives more binding and smaller radius.  Comparing the different $0^+$
and $2^+$ states the tendency is also clearly that the smallest
binding lead to the largest radius. The trends for angular momentum is
that $s$-waves easier extend to larger distances whereas $d$-waves and
higher are confined by the centrifugal barrier.  This is seen for the
$3^+$ state which is weaker bound with smaller angular momenta than
the $4^+$ state, and consequently also significantly larger.

When we assume equal matter and charge radius for the core we find
that the charge radii are slightly larger than the corresponding
matter radii.  This is in spite of the ``natural'' reduction factor of
$9/10$ in Eq.(\ref{e87}).  The reason is found in a core radius which
is substantially smaller than both the proton-proton distance as well
as the distance from their center of mass and the core, see
table~\ref{tab12}.  Combined with a smaller weight in Eq.(\ref{e87})
on these terms for matter compared to charge radii this results in
these larger charge radii.  Recent measurements confirm that the
charge radius is larger than the matter radius \cite{gei08,oza01}. 
If we compare our values with FMD calculations \cite{gei08} we can see that 
our matter radius is bigger and closer to experimental data. On the other hand the FMD charge radius is in better agreement with the 
experimental value.

\begin{table}
\caption{Root mean square (rms) radii in fm for the different bound states
in $^{18}$Ne with the three proton-core interactions.  Charge and
matter radii are shown in first and second row, respectively. They
employ the root-mean square radius of $2.71$~fm for the $^{16}$O
core. The only known experimental values for $^{18}$Ne are the root
mean square charge radius, 2.971$\pm$0.020~fm \cite{gei08}, and matter radius, 2.81$\pm$0.14~fm \cite{oza01}, for
the ground state. In \cite{gei08} we can also find FMD calculations that give 2.93~fm and 2.70~fm for the charge and 
matter radii of the ground state.}
\begin{tabular}{|c|ccc|}
\hline
      & $W_I$ & $W_{II}$ & $W_{III}$    \\ \hline
$0^+_1$ & $2.82$   &  $2.80$   &   $2.74$  \\
         & $2.78$   &  $2.77$   &   $2.73$  \\  \hline
$2^+_1$ & $2.86$   &  $2.84$   &   $2.78$   \\
         & $2.80$   &  $2.79$   &   $2.75$   \\  \hline
$4^+$   & $2.82$   &  $2.83$   &   $2.84$   \\
          & $2.78$   &  $2.78$   &   $2.79$    \\  \hline
$0^+_2$ & $3.08$   &  $3.30$   &   $2.80$  \\
      & $2.93$   &  $3.07$   &   $2.76$  \\  \hline
$2^+_2$ & $2.92$   &  $2.92$   &   $2.86$   \\
      & $2.84$   &  $2.83$   &   $2.80$    \\  \hline
$3^+$   & $3.47$   &  $3.44$   &   $3.30$   \\
      & $3.17$   &  $3.15$   &   $3.06$   \\  \hline
\end{tabular}
\label{tab11}
\end{table}

The average size can be distributed between distances of the different
constituents, i.e. in the present case the proton-core and
proton-proton distances. These root mean square radii for the computed
bound states are given in table~\ref{tab12} for the different
potentials. These two-body distances within the three-body system also
follow the general trends of binding energy and angular momentum. The
distance between the two protons is in all cases larger than the
proton-core distance.  This is because the proton-core attraction is
decisive for the binding of all three-body states.  For the weakly
bound $3^+$ state this difference is substantially larger due to the
small binding energy.  The sizes show that the protons are located
substantially outside the surface of the core. This is obviously
helping to decouple core and valence degrees of freedom, and validate
the model assumptions in the treatment as a three-body system.  In
general the trends from the overall rms radii in table~\ref{tab11} are
maintained.

\begin{table}
\caption{For the different computed states in $^{18}$Ne, and the
different proton-core potentials, root mean square distances (in fm)
$\langle r_{pp} \rangle^{1/2}$ and $\langle r_{cp} \rangle^{1/2}$,
where $p$ and $c$ denote an external proton and the core,
respectively.  }
\begin{tabular}{|cc|ccc|}
\hline
         &                                 & $W_I$ & $W_{II}$
&$W_{III}$   \\ \hline
$0^+_1$ &$\langle r^{2}_{pp} \rangle^{1/2}$ & 4.0   &  3.7   &   3.4        \\
        &$\langle r^{2}_{cp} \rangle^{1/2}$ & 3.4   &  3.3   &   3.0
     \\ \hline
$2^+_1$ &$\langle r^{2}_{pp} \rangle^{1/2}$ & 4.3   &  4.0   &   4.0        \\
        &$\langle r^{2}_{cp} \rangle^{1/2}$ & 3.5   &  3.5   &   3.2
   \\ \hline
$4^+$   &$\langle r^{2}_{pp} \rangle^{1/2}$ & 4.3   &  4.4   &   4.4        \\
        &$\langle r^{2}_{cp} \rangle^{1/2}$ & 3.4   &  3.4   &   3.5
     \\ \hline
$0^+_2$ &$\langle r^{2}_{pp} \rangle^{1/2}$ & 6.0   &  6.6   &   4.5        \\
        &$\langle r^{2}_{cp} \rangle^{1/2}$ & 4.4   &  5.2   &   3.2
     \\ \hline
$2^+_2$ &$\langle r^{2}_{pp} \rangle^{1/2}$ & 4.8   &  4.8   &   4.8        \\
        &$\langle r^{2}_{cp} \rangle^{1/2}$ & 3.8   &  3.8   &   3.5
     \\ \hline
$3^+$   &$\langle r^{2}_{pp} \rangle^{1/2}$ & 8.2   &  8.1   &   7.5        \\
        &$\langle r^{2}_{cp} \rangle^{1/2}$ & 5.8   &  5.7   &   5.2
     \\ \hline
\end{tabular}
\label{tab12}
\end{table}

\begin{figure}
\epsfig{file=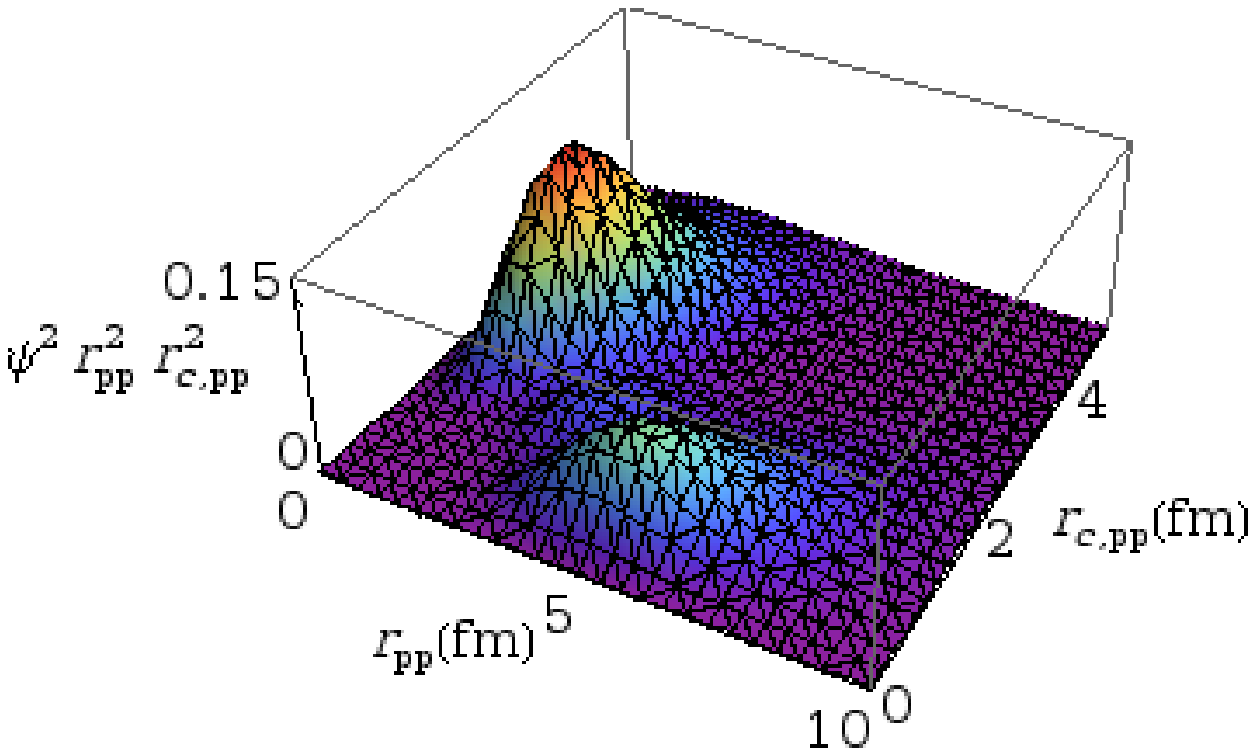,width=8cm,angle=0}
\epsfig{file=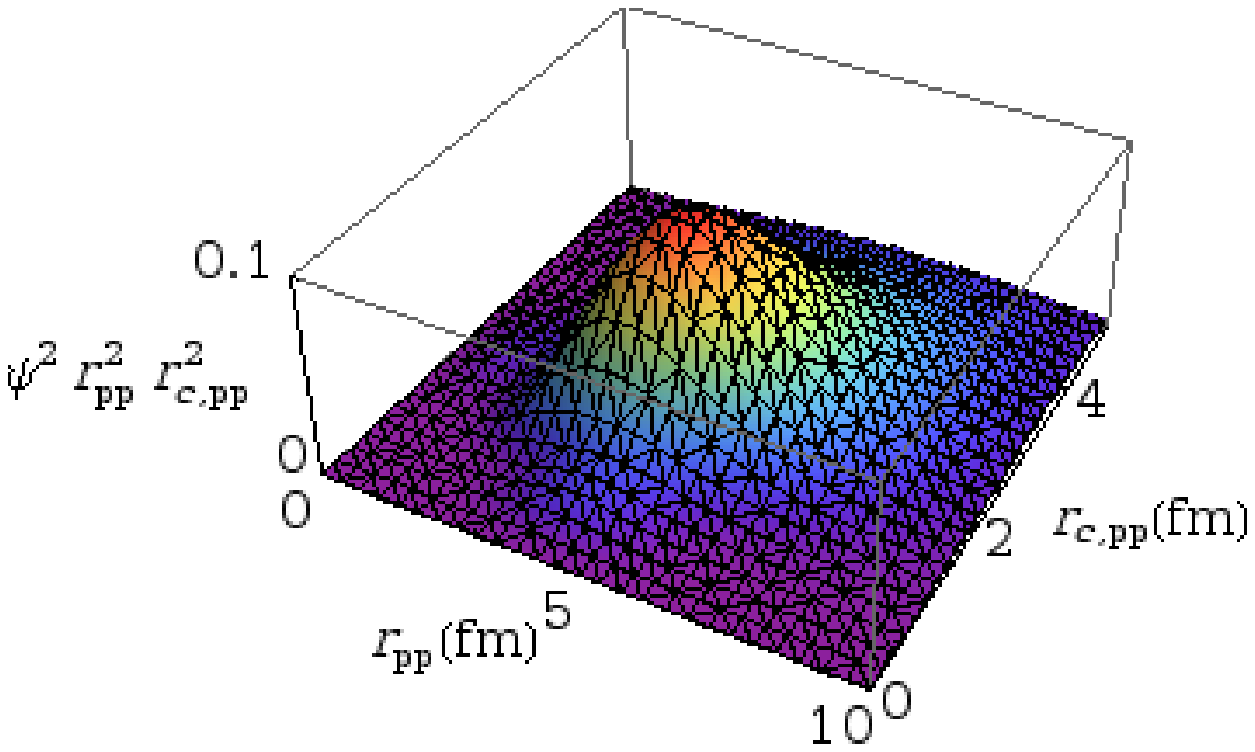,width=8cm,angle=0}
\caption{Contour diagram for the probability distribution of 0$^+_1$ (upper) 
and 0$^+_2$ (lower) states in $^{18}$Ne. The square of the three-body
wave function is integrated over the directions of the two Jacobi
coordinates.}
\label{fig2}
\end{figure}

The average sizes in tables~\ref{tab11} and \ref{tab12} are results of
the probability distributions.  They are shown in Fig.\ref{fig2} for
the two $0^+$ states as functions of the distances between the two
protons and their center-of-mass and the core.  Both distributions
have a tail in the proton-proton distance extending to about
$10$~fm. The fall-off in $r_{c,pp}$ seems to be faster reaching no
more than about $5$~fm.  The ground state has two separated peaks
around the points $(r_{pp},r_{c,pp}) \approx$ $(1.5,2.7),$ $(5.0,0.8)$
(all in fm) where the last is much smaller and somewhat broader.  In
contrast the second $0^+$ state has only one peak at
$(r_{pp},r_{c,pp}) \approx (4.2,2.2)$.  Since the angular momentum
decompositions are rather similar these differences must arise from
the interference between the adiabatic components. The prominent peak
in $0^+_1$ is in $0^+_2$ moved to larger distances between the protons
and the small peak is at the same time moved to somewhat larger
distances in $r_{c,pp}$. The result is that $0^+_2$ has one broad
peak.

The same pattern is found for the two $2^+$-states with two peaks for
$2^+_1$ at $(r_{pp},r_{c,pp}) \approx (1.9,2.7),(5.0,1.0)$ and one
broad peak for $2^+_2$ at $(r_{pp},r_{c,pp}) \approx (5.0,1.8)$.  The
positions are also almost the same as for the $0^+$-states where the
latter position tends to be at smaller distances.  The $4^+$-state has
one peak at $(r_{pp},r_{c,pp}) \approx (3.5,2.0)$ which is an almost
equal sided triangle.  The $3^+$-state has one peak at
$(r_{pp},r_{c,pp}) \approx (4.6,2.2)$.


\subsubsection{Unbound states}

The negative parity states are in principle all resonances but except
for the $1^-_1$ $3^-_2$ states they are all computed as bound states
with respect to the $3^-$ core excitation.  Therefore the radial
moments are for these ($3^-$ based) well defined and together with
partial wave decomposition characteristic for the structures.  In
table~\ref{tab11a} we give root mean square radii of matter and charge
together with distances between protons and core for these
states. Here in order to calculate matter and charge radii we have
used the same radii for the core.
 
\begin{table}
\caption{Root mean square (rms) radii in fm for different negative
parity resonances in $^{18}$Ne with proton-core interactions III (for
the $1^-_1$ state) or IV (for the $1^-_2$, $0^-$, $3^-_1$, and $2^-$ states).
Charge and matter radii are shown in second and third column.  We used
the root-mean square radius, $2.71$~fm, of the $^{16}$O ground state
instead of the unknown value for the $3^-$ excitation.  In the fourth
and fifth column we show root mean square distances (in fm) $\langle
r_{pp} \rangle^{1/2}$ and $\langle r_{cp} \rangle^{1/2}$, where $p$
and $c$ denote an external proton and the core, respectively. }
\begin{tabular}{|c|cccc|}
\hline
 & $\langle r_c^2 \rangle^{1/2}$  & $\langle r^2 \rangle^{1/2}$ &
 $\langle r_{pp} \rangle^{1/2}$  &  $\langle r_{cp} \rangle^{1/2}$ \\
\hline
$1^-_1$ & $3.00$ & $2.88$   &  $5.80$ &  $4.12$  \\
$1^-_2$ & $2.86$ & $2.80$   &  $4.36$ &  $3.26$    \\
$0^-$   & $2.77$ & $2.75$   &  $4.11$ &  $3.10$    \\
$3^-_1$   & $2.72$ & $2.72$   & $3.64$  &  $2.88$   \\
$2^-_1$ & $3.01$ & $2.90$   &  $4.37$ &  $3.87$  \\  \hline
\end{tabular}
\label{tab11a}
\end{table}

As for the positive parity bound states the charge radii are usually
slightly larger than matter radii, see table~\ref{tab11a}. In general
the positive and negative parity states are almost of the same size
even though the binding to the $3^-$ core is larger by several
MeV. Also the internal distances remain essentially the same with the
proton distance as the largest. Not surprisingly the tendencies with
binding energy and angular momentum follow the general rules explained
for the positive parity states. Only the $1^-_1$ state of roughly zero
energy has a sufficiently well defined radial structure to be included
in table~\ref{tab11a}.

\begin{figure}
\epsfig{file=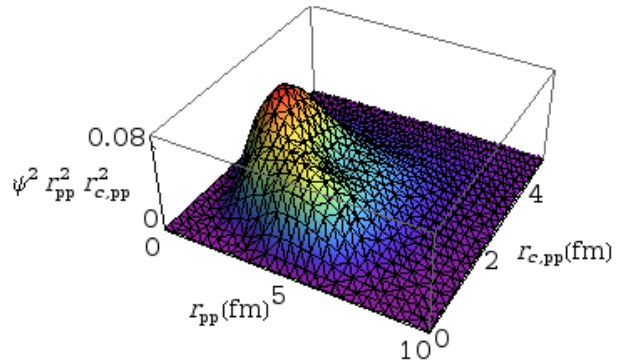,width=8cm,angle=0}
\caption{Contour diagram for the probability distribution of 
2$^-_1$ 
 state in $^{18}$Ne. The square of
the three-body wave function is integrated over the directions of the
two Jacobi coordinates.}
\label{fig3a}
\end{figure}

The probability distributions are all rather similar.  Both
$1^-$-states have only one well-defined peak at $(r_{pp},r_{c,pp})$
$\approx (1.9,2.7),(3.5,1.8)$.  The first of these resemble the lowest
$0^+$ and $2^+$state and the second more resembles the $4^+$ state. We
also find one peak for both $0^-$, $2^-_1$ and $3^-_1$-states at
$(r_{pp},r_{c,pp}) \approx$ $(3.1,1.6),(1.9,2.2),$ $(2.3,1.6)$. For
$2^-_1$ a second peak is indicated at the side of the main peak
resulting in an intermediate structure between the two we have already
seen for bound states (see Fig. \ref{fig3a}). The $0^-$ state resemble
the $4^+$ state with the two protons close to the core but rather far
from each other.  

The probability distributions for the resonances built on the $0^+$
core ground state are not well defined since their energies are above
threshold and the radial wavefunctions therefore spread out to
infinitely large distances. However, they all have large amplitudes
around the minimum in the corresponding adiabatic potentials.

\subsection{Electromagnetic decays}

The bound states can only decay electromagnetically.  The
corresponding observable transition probabilities are critically
depending on the structures. Thus they provide experimental tests and
we therefore compute the lifetimes for future comparison. The
three-body states below $6.13$~MeV built on the core-excited $3^-$
state are also bound states. They can therefore only decay by
$\gamma$-emission to lower lying states either by maintaining the same
cluster structure or by $E3$-decay of the core excited state. In both
cases we can compute the electromagnetic transition probabilities.
The selection rules determine the dominating transitions which can be
of both electric and magnetic origin.

The effective charge of the proton $e_{eff}$ is for these estimates
determined by renormalizing the calculated $E2$-transition strength
from first excited to ground state in $^{17}$F to the measured value
of 25 Weisskopf units or $62.96$e$^2$~fm$^4$ \cite{til95}. This gives a value
amazingly close to unity $e_{eff}/|e| = 0.99$ which is used in the
three-body computations.

The structures are sufficiently similar for the different interactions
to allow estimates with only one potential for each state. We choose
potential I and IV for positive and negative parity states,
respectively.  The selection rules determine the dominating
transitions which can be of both electric and magnetic origin.
Transitions within the same parity are then dominated by $E2$ or $M1$
emissions, and between different parity states predominantly by $E1$.
If $E1$ is forbidden the much weaker $M2$ or $E3$ transitions may
determine the lifetime, but to this level of accuracy the neglected
mixture of ground and excited core-states could contribute.

\subsubsection{Multipole operators}

The electric multiple operators are defined as:
\begin{equation}
{\cal M}_\mu(E\lambda)=e\sum_{i=1}^A Z_i r_i^\lambda Y_{\lambda,\mu}(\hat{r}_i)
\label{eq6}
\end{equation}
where $A$ is the number of constituents in the system, each of them
with charge $eZ_i$, and where $\bm{r}_i$ is the coordinate of each of
them relative to the $A$-body center of mass.

The electric multipole strength functions are defined as:
\begin{eqnarray}
{\cal B}(E\lambda, I_i\rightarrow I_f) &=&
\sum_{\mu M_f} |\langle I_f M_f|{\cal M}_\mu(E\lambda) | I_i M_i \rangle|^2
  \nonumber \\ & = & \frac{1}{2I_i+1} |\langle I_f ||{\cal M}(E\lambda) || I_i \rangle|^2
\label{eq7}
\end{eqnarray}

The ${\cal M}_\mu(E\lambda)$ operator in Eq.(\ref{eq6}) can then be
rewritten as:
\begin{eqnarray}
{\cal M}_\mu(E\lambda)& = &
 e\sum_{i=1}^{A-2} Z_i |\bm{r}_c+\bm{r}_i^\prime|^\lambda Y_{\lambda,\mu}(\widehat{\bm{r}_c+\bm{r}_i^\prime})
   \nonumber \\
&    + & e\sum_{j=1}^2 Z_j r_j^\lambda Y_{\lambda,\mu}(\hat{r}_j)
\label{eqa2}
\end{eqnarray}
where the index $i$ runs over the $A-2$ constituents in the core, and
$j$ labels the two external nucleons.

As in \cite{rom08} we can rewrite into intrinsic core and external
valence nucleon coordinates. The results turn out to have the form
\begin{eqnarray}
&&{\cal M}_\mu(E\lambda)= e \sum_{i=1}^{3} Z_i 
 r^{\lambda}_i Y_{\lambda,\mu}(\hat{r}_i)+{\cal M}_\mu(E\lambda,c)
\label{eqa4}
   \\ \nonumber
&  + & \sum_{k=1}^{\lambda-1}\sum_{m=-k}^{k} f_{\lambda}(k,m,\mu) 
 r_c^{k}  Y_{k,m}(\hat{r}_c) 
 {\cal M}_{\mu-m}(E(\lambda-k),c)
\end{eqnarray}
where the two first terms refer to independent valence and core
degrees of freedom, respectively.  The last terms describe
simultaneous transitions of core and valence particles where
$f(m,\mu)$ is a well defined function of its indices. Then, since the
only allowed core transition is ${\cal M}_\mu(E3)$, only the second
term contributes to transitions between the two different core states.
For transitions between the same core state the mixed terms may in
principle contribute. However, these terms are accounted for by the
effective charge of the proton which was adjusted to describe the
$E2$-transition in $^{17}$F.  Thus also these terms should not be
included.

The magnetic multipole operator, ${\cal M}_\mu(M\lambda)$, has the
opposite parity of ${\cal M}_\mu(E\lambda)$ and give rise to much
smaller rates for the same $\lambda$.  Thus ${\cal M}_\mu(M2)$ only is
active between negative parity states where the $E3$ core transition
is necessary in the present cases. The $(M2)$ transitions are then
forbidden. On the other hand $(M1)$ conserves parity, allows unchanged
core structure, and may compete with $(E2)$ transitions. We shall
therefore only consider the $(M1)$ operator which is defined as:
\begin{eqnarray}   \label{e20}
 {\cal M}_\mu(M1) = \frac{e\hbar}{2Mc}\sqrt{\frac{3}{4\pi}} 
\sum_i (g^{(i)}_s \vec s_i  + g^{(i)}_\ell \vec \ell_i)_{\mu}  \;,
\end{eqnarray}
where $\mu$ labels the spherical component of an operator, and the
constants $g_s$ and $g_\ell$ depend on the constituent particles
$i$. The magnetic multipole transition strength is defined as for the
electric case (see Eq.(\ref{eq7})).

The core has a charge of $8$ units suggesting that $g_\ell^{(c)}=8$.
For the positive parity states where the core angular momentum is zero
we use a vanishing effective spin $g$-factor, i.e. $g_s^{(c)}=0$.  For
the negative parity states where the core angular momentum is $3$ the
$g$-factor is unknown but also of little interest since the dominating
decay probabilities are determined by other transitions.  We use the
free proton value of $g_s^{(p)}$=$5.586$ and we use again an effective
proton charge $g_\ell^{(p)}$=$0.99$.  The relevant transition
operators are then defined and we can compute the observable
transition strengths.

\subsubsection{Transition strengths}

The ${\cal B}(E2)$-transition for $^{17}$F from $1/2$ to $5/2$ amounts to about
$63$e$^2$fm$^4$ corresponding to $25$~W.u. and a width of $1.6\times
10^{-6}$~eV \cite{til95}. This is within $1\%$ the same as computed
from the relative two-body wavefunctions.  This indicates that an
effective charge of $1$ should be used for the low-energy sequence of
states in $^{18}$Ne which all are dominated by $E2$ transitions.  We
collect in table~\ref{tab16} the four possible ${\cal B}(E2)$ values
from first and second $2^+$ to the other positive low-lying parity
states.  As usual the reduced matrix element should be multiplied by
$1/(2J_i+1)$ depending on initial and final states in the transition.
The table values include this factor and reflect the chosen direction
of the transition.

\begin{table}
\caption{$E2$-transition strengths from first and second $2^+$ states to 
neighboring $J^{+}$ states expressed in units of $e^2$~fm$^4$. The
first two rows (upper part) contain measured values from \cite{til95}. The next two rows (central part)
correspond to our calculation. The last two rows (lower part) show computed values obtained in \cite{duf04}.} 
\begin{tabular}{|c|cccc|}
\hline
 ${\cal B}$ &  $0_1^+$ & $0_2^+$ & $2_i^+$ & $4^+$ \\ \hline

 ${\cal B}(E2,2^+_1\rightarrow J^+)$ & $48(5)$ & $2.7(13)$ & $-$ & $43(6)$ \\
 ${\cal B}(E2,2^+_2\rightarrow J^+)$ & $1.8(9)$ & $-$ & $-$ & $-$\\ \hline
 ${\cal B}(E2,2^+_1\rightarrow J^+)$ & $21.53$ & $2.199$ & $2.445$ & $31.71$ \\
 ${\cal B}(E2,2^+_2\rightarrow J^+)$ & $0.322$ & $7.154$ & $2.445$ & $1.179$ \\ \hline
 ${\cal B}(E2,2^+_1\rightarrow J^+)$ & $32.09$ & $-$ & $-$ & $59.26$ \\
 ${\cal B}(E2,2^+_2\rightarrow J^+)$ & $0.084$ & $-$ & $-$ & $-$ \\ \hline
\end{tabular}
\label{tab16}
\end{table}

For potential I the computed table values (central part of table \ref{tab16}) are all systematically
smaller than the measured results (upper part of table \ref{tab16}), i.e. smaller by factors of 2.2,
1.22, 1.36, respectively for the known $2^+_1$ transitions in the
third row of table~\ref{tab16}.  For ${\cal B}(E2,2^+_1\rightarrow
0_1^+)$ there are newer measurements resulting in $23\pm4e^{2}fm^{4}$
and $27\pm4e^{2}fm^{4}$ in~\cite{ril00} which are in better agreement
with our values. However, later on the same author published
in~\cite{ril03} the results of a measurement of the life time of this
excited $2^+_1$ state which is more consistent with the value we used
from \cite{til95}.  The computed $2^+_2\rightarrow 0_1^+$ transition
is smaller by a factor 5.5 but on top of the varying experimental
results the measured value in \cite{til95} is given with a rather
large uncertainty which could reduce this discrepancy to a factor of
2.7.  The discrepancies are reduced if we correct all numbers
by scaling the root mean square value of the ground state from the calculated
$2.78$~fm to the measured value of $2.81$~fm as found in
\cite{oza01}.

The remaining deviations are now within acceptable ranges for a model
where core polarization is neglected.  The largest discrepancy appears
for the $0^+$ state which might have the strongest influence from the
lowest core excitation of the same quantum number, $0^+$.  The
phenomenological procedure to correct for that effect is to use an
effective charge larger than unity which then accounts for influence
beyond the single-particle degrees of freedom.  Most of the
transitions are substantially larger than corresponding to one single
particle unit.  This usually is a signal of the need for an effective
charge larger than unity which in turn implies that core degrees of
freedom are important.  On the other hand the transition probabilities
are not consistent with collective vibrational motion as the spectrum
of excited states otherwise could indicate. Furthermore, the main
variation is picked up by the three-body model.

With the radius of $1.2 A^{1/3}$~fm the single particle Weisskopf unit
for ${\cal B}(E2)$ is 2.690$e^2$fm$^4$ which indicates that the
proton-core $E2$-transition of about $25$~W.u. is constructed by
substantially more than a simple single-particle transition.  On the
other hand this $E2$ transition in $^{17}$F is reproduced with the
proper wavefunctions and an effective proton charge of one.  Then it
is not unreasonable to expect that the three-body system should be
approximately describable without active core degrees of freedom as
well or perhaps rather with appropriate effective charges from the
two-body subsystem.

The transition probability from $2_2^+$ to $2_1^+$ receives also a
contribution from $M1$.  We find ${\cal B}(M1,2^+_2\rightarrow 2_1^+)
= 9.7\times10^{-4}$~$e^2$fm$^2$ 
which is smaller than the
measured value of $0.0017\pm0.0007$~$e^2$fm$^2$ (0.088$\pm$0.038 W.u.) corresponding to a
width of $9.5\times10^{-3}$~eV \cite{til95}.  The small computed value
has the inherent uncertainty arising from spin polarization which can
lead to a substantial correction.  In any case this decay seems to be
dominated by $M1$.

\begin{table}
\caption{$E2$ and $M1$-transition strengths from the second $3^+$ state 
to lower-lying $J^{+}$ states expressed in units of $e^2$~fm$^4$ and
$e^2 fm^2$, respectively. The third row gives the ratio of partial
widths for the two decay modes. The fourth and fifth rows shows the values from \cite{duf04} }
\begin{tabular}{|c|ccc|}
\hline
 ${\cal B}$ &  $2_1^+$ & $4^+$ & $2_2^+$  \\ \hline
 ${\cal B}(E2,3^+\rightarrow J^+)$ & $0.0596$ & $24.0$ & $4.27 $  \\
 ${\cal B}(M1,3^+\rightarrow J^+)$ & $0.0240$ & $1.76\times10^{-7}$ & $1.16\times10^{-3}$  \\ 
 $\Gamma_{E2}/\Gamma_{M1}$ & $1.35\times10^{-5}$ & 147 & 0.00250  \\ \hline
 ${\cal B}(E2,3^+\rightarrow J^+)$ & $0.107$ & $5.89$ & $14.4 $  \\
 ${\cal B}(M1,3^+\rightarrow J^+)$ & $1.54\times10^{-3}$ & $3.84\times10^{-3}$ & $1.35\times10^{-4}$   \\ \hline
\end{tabular}
\label{tab17}
\end{table}

The decay modes of lowest multipolarity for the $3^+$ state are $E2$
and $M1$ where the final state can be any of the three states shown in
table~\ref{tab17}.  The corresponding decay widths are given by
$\Gamma_{E2} = 4\pi/75 (E_{\gamma}/\hbar c)^5 {\cal B}(E2) $ and
$\Gamma_{M1} = 16\pi/9 (E_{\gamma}/\hbar c)^3 {\cal B}(M1)$. The ratio
of widths is then $\Gamma_{E2}/\Gamma_{M1} = 0.03 (E_{\gamma}/\hbar c)^2 {\cal
B}(E2) / {\cal B}(M1)$.

The $E2$ transition from $3^+$ is dominated by decay to the $4^+$
state.  The $E2$ transitions to the two $2^+$ states are smaller by
factors of about $6$ and 30, respectively.  These decay probabilities
are essentially completely arising from single-particle proton
transitions between orbits around the $^{17}$F structure. The relative
sizes correspond directly to the probabilities of the largest $E2$
allowed configurations in tables~\ref{tab4}, \ref{tab5} and \ref{tab6}.

The $M1$ transitions are also essentially due to proton transitions
between orbits around the $^{17}$F structure.  They are all very small
but still varying by orders of magnitude. The sizes strongly indicate
that the dominating parts of these transitions are forbidden by $M1$
selection rules due to the single particle character of the operator.
The decisive selection rules are $\Delta\ell_x=\Delta\ell_y=0$, and
$\Delta s_x=\Delta s_y=0$.  For the $\ell$-part also $\Delta S=0$ and
analogously $\Delta L=0$ for the $s$-part.  The $3^+$ state is from
table~\ref{tab6} seen to be dominated by
$(\ell_x,\ell_y,L,s_x,S)=(2,0,2,1/2,1),(0,2,2,1/2,1)$ in the second
and third Jacobi coordinate system. The selection rules then prohibits
$M1$ transitions to the $4^+$ state between the dominating components,
see table~\ref{tab5}, while much smaller components still contribute
resulting in the small value in table~\ref{tab17}.  The transitions to
the $2^+$ states can only proceed via small components, see
table~\ref{tab4}, and the contributions in table~\ref{tab17} are
consistently also very small.

In any case the small $M1$ transition probabilities cannot be precise
since admixtures of various kinds or changes of the already
contributing components could change the numerical results completely.
Assuming the values in table~\ref{tab17} it is interesting to see that
decays into $2_{1}^+$ and $2^+_2$ are dominated by $M1$ while
the $4^+$ state preferentially would be reached by $E2$.

In the $M1$ computations the effective values of the $g$-factors are
rather uncertain and polarization effects could change them.  The
$M1$-transitions are particularly sensitive to spin polarization which
is able to change $g_s^{(p)}$ significantly. The orbital $g$-factors
can be expected to have less uncertainty on a level similar to the
effective charges.  Unfortunately, no values are available from the
two-body subsystem which otherwise could have provided the basis for
comparing two and three-body effective $g$-factors. More precisely, in
this way be able to give an answer to which degree of adjustment to
two-body data is necessary to reproduce the three-body transition
properties. 

The cluster model in \cite{duf04} has a very different structure with
only two-body components, either $\alpha + ^{14}$O or $p + ^{17}$F.  A
number of transitions between positive parity states are computed.  We
give the transition strengths in the caption of table \ref{tab16} and in table
\ref{tab17}.  The ${\cal B}(E2,2^+_1\rightarrow 0^+_1)$ in \cite{duf04} is between our value and
measurement, ${\cal B}(E2,2^+_1\rightarrow 4^+)$ is larger than measured where our
result is smaller, ${\cal B}(E2,2^+_2\rightarrow 0^+_1)$ in ref \cite{duf04} deviate from
experiment by a factor of 20 compared to our factor of about 5.  The
transition ${\cal B}(M1,2^+_2\rightarrow 0^+_1)$  is also further away from
measurements than ours but both values are very small.  The transition
from the $3^+$ state is the focus of the two-body cluster model in
\cite{duf04}. No experimental values exist. For both E2 and M1 transitions
to the $2^+$ states, \cite{duf04} obtains larger values than in the present
work and vice versa for the $4^+$ state.

The negative parity states are more limited in their decay modes
because they have to change three units of angular momentum on the
core. This means that the transition has to be of order $E3$ or
higher. The only contribution is then seen from Eq.(\ref{eqa4}) to be
accompanied by the core-transition with ${\cal B}(E3) = 259.5
e^2$~fm$^6$ (13.5 W.u.) corresponding to a width of
$2.60\times10^{-5}$~eV.  This value has to be multiplied by the
overlap of the valence wavefunctions.  Therefore the results are
proportional to the overlaps discussed in subsection \ref{sec3c2} but
in any case very small.  Furthermore, any admixture of a $0^+$
component in the dominating $3^-$ core structure of these resonances
would determine the decay probabilities.

These decays would proceed through the small admixtures of core-state
$0^+$ in the negative parity states or by the presumably much smaller
core-state $3^-$ admixture in the positive parity states.  Let us
assume a small mixing amplitude of $\epsilon$ of core-state $0^+$.
Then the decay could be of much smaller multipolarity and the rate
therefore much larger.  If $E1$ is allowed it would dominate but
should be reduced by the amplitude $\epsilon$.  Thus the $E3$ core
transition from $J^-$ has to compete with the $E1$ rate related to the
matrix element: 
\begin{equation}
\langle 0^+(core);J^+=|J^- \pm 1||{\cal
M}(E1)|0^+(core);J^- \rangle. 
\end{equation}
The $E1$ rates should then be
multiplied by $\epsilon^2$ indicating that even with a $0.1\%$
admixture the $E1$-transition would dominate.

\section{Summary and conclusions}
\label{sec7}
The properties of $^{18}$Ne have been investigated assuming a
three-body structure with an $^{16}$O core and two protons.  The aim
is to establish these structures which are necessary ingredients in
computations of momenta of protons and $^{16}$O from two-proton
decays.  We use the hyperspherical adiabatic expansion method for the
Faddeev equations.  The proton-proton interaction reproduce low-energy
scattering data.  The proton-core two-body interaction is adjusted to
reproduce the bound $s$ and $d$-states of the proton-$^{16}$O
($^{17}$F) subsystem. The negative parity two-body $^{17}$F states are
unbound but may still be used to constrain the $p$-wave interaction.
The $p$-waves are then treated in different ways, i.e. first by using
the same potential as for $s$ and $d$-waves and second by adjusting
independently to the measured levels.  Three different potentials are
constructed with the ground state structure of the $0^+$ structure for
$^{16}$O with different prescription to account for the Pauli blocking
of the occupied core states.

The unbound negative parity states are relatively high-lying. They
appear to be more related to core excited $^{16}$O states than to
single-particle $p$-waves.  We then use the $^{16}$O $3^-$ excited
state as building block.  The sequence and spacing of the low-lying
$^{17}$F negative parity resonances support this interpretation as
essentially uncoupled structures.  The corresponding interactions are
adjusted to reproduce the lowest-lying negative parity states of
$^{17}$F with an inert $3^-$ excited $^{16}$O.  We follow this
conjecture and exploit the consequence of uncoupled three-body
structures built on these different $^{16}$O structures.

The first results are energies and structures of the five lowest-lying
bound states of $^{18}$Ne.  The energies are reproduced with the usual
accuracy in such three-body computations of about $200$~keV at least for our potential I and omitting the second 0$^{+}$ state 
(around $500$~keV).  We do
not employ phenomenological three-body potentials to fine tune these
energies.  This is only necessary when precise decay properties are
requested.  The second series of results are bound excited states
built on the $3^-$ core excited $^{16}$O state and decoupled from the
first series of three-body states.  The third series are higher-lying
states arising from the $0^+$ structure of $^{16}$O but for potentials
which only are adjusted to proton-core $s$ and $d$-waves. The
$p$-waves are present and contribute without the relatively strong
attraction necessary to reproduce the $p$-resonance.  These states
have energies below a few MeV above threshold where several states are
known in the measured spectrum.  Furthermore in most cases these states
are separated by several MeV from the states of the same angular
momentum and parity arising from the $3^-$ core excited $^{16}$O
state.  These structures may therefore still be essentially uncoupled.

The detailed structures of all these states are extracted and
discussed in terms of the partial waves of the subsystems building
these three-body states. These predictions would be ingredients in
future tests involving transfer into these states or breakup or decay
from them. We also provide root-mean square radii for all states for
both matter and charge and division into proton-proton and proton-core
distances.  Not surprisingly the distributions are all peaked at
distances corresponding to the minima in the adiabatic potentials. In
some cases also a smaller peak appear nearby due to different
contributing configurations.

The transitions between these states are rather sensitive to their
electromagnetic structures. For the bound states E2 transitions
dominate and we compute them and compare if possible to measured
values. In all cases we find very large decay rates although
systematically lower than experiments by a factor between 1.5 and 2. However in the $2^+_2\rightarrow 0_1^+$ transition we have
a much larger difference with the experimental
results although this value has a
large uncertainty, as already mentioned, which could reduce this difference. 
The M1 transitions between bound states only contribute between the
two $2^+$ states, and although almost forbidden also between the $3^+$
and the $2^+,4^+$ states. The $M1$ transition rate between $2^+$
states is an order of magnitude smaller than the measured value.  The
widths for transitions between $2^+$ states and from $3^+$ to the
$2^+_1$ and $2^+_2$ states are dominated by $M1$ while $3^+$ to $4^+$
is dominated by $E2$.

We compare to values from a two-body cluster model with very different
structures of all the investigated states. For the lowest positive
parity $0,2,4$ states the $E2$-transitions are in both models within
factors of about 2 from experiments. The transition from second
$2^+$ to ground state is improved by a factor of 4 in our model,
still missing about a factor of 5 compared to the measurement.
Transitions involving the $3^+$ state are very different in the two
models, and here especially the very small M1-values.

Electromagnetic Transitions between states related to different core
structures are in the schematic model determined by the E3-transitions
between the core states. However, in reality even very small
admixtures of $0^+$ core structure in the wavefunctions of dominating
$3^-$ core structure could easily enhance these transitions by many
orders of magnitude by allowing lower multipolarity. Furthermore, such
small admixtures would lead to three-body decay widths of these
resonances completely dominating over the electromagnetic decays.
These decays would then proceed through the small component of the
$0^+$ core structure and decisive as determined only by the strong
interaction.

In summary, we have in details investigated the three-body structure
of the low-lying states of $^{18}$Ne.  The bound state energies are
rather well reproduced by properties of the two-body subsystems. The
resonances can not be compared directly to measured values due to lack
of detailed information. These resonances fall in two groups related
to two different structures ($0^+$ and $3^-$) of the $^{16}$O
core. The small experimental widths of many of the $^{18}$Ne negative
parity resonances can be explained by their main structure as bound
states with an $^{16}$O excited $3^-$ state. Their decay is then
through small admixtures of the $^{16}$O ground state in their
wavefunctions.  The E2-transitions between bound states are
systematically smaller than the measured values but within the range
of moderate values of effective charges and $g$-factors.  These
investigations are necessary for computations of three-body decays of
$^{18}$Ne resonances but useful directly by providing detailed
information about the structure of low-lying $^{18}$Ne states.

\begin{center}
{\bf ACKNOWLEDGMENTS}
\end{center}
This work was partly supported by funds provided by the DGI of MEC
(Spain) under Contract No. FIS2008-01301.  J.A.L. acknowledges a
Ph.D. grant from the University of Seville and C.R.R. acknowledges
support by a predoctoral I3P grant from CSIC and the European Social
Fund.

\appendix

\begin{center}
{\bf APPENDIX}
\end{center}

\section{Wavefunctions Components}

We give the contributions larger than $0.1\%$ of each Faddeev
component for a number of computed states as discussed in section
\ref{sec3}.  Results for both the two different Jacobi coordinates are 
given for each state.

\begin{table} 
\caption{The same as table \ref{tab3} for the 2$^+$ states in $^{18}$Ne. }
\begin{tabular}{|cccccc|ccc|}
\hline
$\ell_x$ & $\ell_y$ & $L$ & $s_x$ & $S$ & $K_{max}$ & $W_I$ & $W_{II}$
& $W_{III}$ \\
\hline
   2  &  0  &  2  &  0  &  0  &  90 &  27.6   &  22.1  &  25.8  \\
      &     &     &     &    &      &   9.9   &   9.4  &  14.4  \\
   1  &  1  &  1  &  1  &  1  &  90 &  11.8   &  10.9  &   5.5 \\
      &     &     &     &    &     &   27.1   &  27.8  &  31.4 \\
   1  &  1  &  2  &  1  &  1  &  80 &   8.2   &   8.0  &  12.9 \\
      &     &     &     &    &     &    9.4   &   9.1  &   8.1 \\
   0  &  2  &  2  &  0  &  0  & 120 &  51.4   &  58.2  &  54.9 \\
      &     &     &     &    &     &   13.0   &  13.4  &  16.8 \\
   2  &  2  &  2  &  0  &  0  & 100 &   1.0   &   0.9  &   0.8 \\
      &     &     &     &    &     &   40.7   &  40.3  &  29.2 \\ \hline
   1  &  1  &  2  & 1/2 &  0  &  70 &   2.6   &   7.2  &   2.8 \\
      &     &     &     &    &     &    0.1   &   0.2  &   0.1 \\
   1  &  1  &  2  & 1/2 &  1  &  60 &   1.3   &   0.1  &   0.1 \\
      &     &     &     &    &     &    0.1   &   0.1  &   0.1 \\
   2  &  2  &  1  & 1/2 &  1  & 100 &  10.7   &   9.7  &   5.3 \\
      &     &     &     &    &     &   22.8   &  23.5  &  28.6 \\
   2  &  2  &  2  & 1/2 &  0  & 120 &  25.5   &  23.3  &  15.7 \\
      &     &     &     &    &     &   31.4   &  32.1  &  40.5 \\
   2  &  2  &  3  & 1/2 &  1  &  80 &   3.2   &   2.9  &   1.9 \\
      &     &     &     &    &     &    4.9   &   5.1  &   6.1 \\
   2  &  0  &  2  & 1/2 &  0  & 100 &  22.3   &  22.3  &  30.2  \\
      &     &     &     &    &     &   12.9   &  12.3  &   7.8 \\
   2  &  0  &  2  & 1/2 &  1  &  80 &   7.1   &   6.7  &   7.6 \\
      &     &     &     &    &     &    7.6   &   7.3  &   4.4 \\
   0  &  2  &  2  & 1/2 &  0  & 100 &  21.5   &  20.7  &  28.9 \\
      &     &     &     &    &     &   12.5   &  12.0  &   7.7 \\
   0  &  2  &  2  & 1/2 &  1  &  80 &   6.9   &   6.7  &   7.4 \\
      &     &     &     &    &     &    7.4   &   7.1  &   4.3 \\
   1  &  1  &  1  & 1/2 &  1  &  60 &   0.1   &   0.3  &   0.0 \\
      &     &     &     &    &     &    0.3   &   0.3  &   0.3 \\
\hline
\end{tabular}
\label{tab4}
\end{table}

\begin{table} 
\caption{The same as table \ref{tab3} for the 4$^+$ state in $^{18}$Ne. }
\begin{tabular}{|cccccc|ccc|}
\hline
$\ell_x$ & $\ell_y$ & $L$ & $s_x$ & $S$ & $K_{max}$ & $W_I$  & $W_{II}$ & $W_{III}$   \\ \hline
   0  &  4  &  4  &  0  &  0  &  80 &  16.6  &  17.7  &  17.3   \\
   1  &  3  &  3  &  1  &  1  & 100 &  34.0  &  32.7  &  33.3    \\
   2  &  2  &  4  &  0  &  0  &  60 &   7.2  &   7.2  &   7.3  \\ 
   3  &  1  &  3  &  1  &  1  & 100 &  32.7  &  33.4  &  32.4   \\
   4  &  0  &  4  &  0  &  0  &  60 &   9.4  &   9.0  &   9.7   \\  \hline
   4  &  0  &  4  & 1/2 &  0  &  40 &   0.1  &   0.1  &   0.1   \\
   3  &  1  &  4  & 1/2 &  0  &  60 &   0.9  &   1.0  &   0.9   \\
   3  &  1  &  3  & 1/2 &  1  &  60 &   0.2  &   0.2  &   0.2    \\
   1  &  3  &  4  & 1/2 &  0  &  40 &   0.0  &   0.1  &   0.1   \\
   1  &  3  &  3  & 1/2 &  1  &  60 &   0.1  &   0.2  &   0.2   \\
   2  &  2  &  4  & 1/2 &  0  & 100 &  31.8  &  32.1  &  32.6    \\
   2  &  2  &  4  & 1/2 &  1  &  60 &   0.1  &   0.1  &   0.1    \\
   2  &  2  &  3  & 1/2 &  1  & 120 &  66.7  &  66.2  &  65.9    \\

\hline
\end{tabular}
\label{tab5}
\end{table}

\begin{table}
\caption{The same as table \ref{tab3} for the 3$^+$ state in $^{18}$Ne. }
\begin{tabular}{|cccccc|ccc|}
\hline
$\ell_x$ & $\ell_y$ & $L$ & $s_x$ & $S$ & $K_{max}$ & $W_I$  &
$W_{II}$ & $W_{III}$   \\ \hline
   1  &  1  &  2  &  1  &  1  & 120 &  54.2 & 54.1  &  77.9     \\
   1  &  3  &  2  &  1  &  1  & 120 &  19.4 & 19.2  &   8.2     \\
   1  &  3  &  3  &  1  &  1  &  40 &   0.0 &  0.1  &   0.0   \\
   3  &  1  &  2  &  1  &  1  & 120 &  19.8 & 20.2  &   8.2   \\
   3  &  1  &  3  &  1  &  1  &  40 &   0.0 &  0.1  &   0.0  \\
   3  &  3  &  2  &  1  &  1  & 100 &   6.6 &  6.4  &   5.7   \\ \hline
   1  &  1  &  2  &  1  &  1  & 100 &   1.2 &  1.1  &   1.2     \\
   1  &  3  &  2  & 1/2 &  1  &  40 &   1.8 &  1.8  &   1.4   \\
   3  &  1  &  2  & 1/2 &  1  &  40 &   0.0 &  0.1  &   0.0   \\
   2  &  0  &  2  & 1/2 &  1  & 120 &  50.1 & 50.0  &  49.9   \\
   0  &  2  &  2  & 1/2 &  1  & 120 &  46.8 & 40.9  &  47.4    \\
   2  &  2  &  2  & 1/2 &  1  &  40 &   0.1 &  0.1  &   0.1   \\ \hline
\end{tabular}
\label{tab6}
\end{table}

\begin{table}
\caption{The same as table \ref{tab3} for the 0$^-$ state in $^{18}$Ne
with the 3$^{-}$ core excited state.  }
\begin{tabular}{|cccccc|c|}
\hline
$\ell_x$ & $\ell_y$ & $L$ & $s_x$ & $S$ & $K_{max}$ & $W_{IV}$  \\ \hline
   1  &  1  &  2  &  1  &  2  &  100 &    83.1     \\
   1  &  3  &  2  &  1  &  2  &   80 &     7.7     \\
   1  &  3  &  3  &  1  &  3  &   60 &     0.1     \\
   3  &  1  &  2  &  1  &  2  &   80 &     7.7     \\
   3  &  1  &  3  &  1  &  3  &   60 &     0.1     \\  \hline
   1  &  1  &  2  &  5/2  &  2  &   60 &    0.1     \\
   2  &  2  &  3  &  5/2  &  3  &   80 &    2.3     \\
   2  &  0  &  2  &  5/2  &  2  &  100 &   49.1      \\
   0  &  2  &  2  &  5/2  &  2  &  100 &   48.3     \\
   1  &  3  &  2  &  5/2  &  2  &   60 &    0.1     \\
   3  &  1  &  2  &  5/2  &  2  &   60 &    0.1     \\ \hline
\end{tabular}
\label{tab8}
\end{table}

\begin{table}
\caption{The same as table \ref{tab3} for the first 3$^-_1$ state 
in $^{18}$Ne built on the 3$^-$ core state.  }
\begin{tabular}{|cccccc|c|}
\hline
$\ell_x$ & $\ell_y$ & $L$ & $s_x$ & $S$ & $K_{max}$ & $W_{IV}$  \\ \hline
  0  &  0  &  0  &  0  &  3  &  120 &   97.4      \\
  1  &  1  &  0  &  1  &  3  &  100 &     1.5     \\
  1  &  1  &  1  &  1  &  2  &   60 &    0.1     \\
  2  &  2  &  0  &  0  &  3  &  100 &    1.0     \\ \hline
  0  &  0  &  0  &  5/2  &  3  &  100 &  42.0       \\
  0  &  0  &  0  &  7/2  &  3  &  100 &  55.7       \\
  1  &  1  &  0  &  5/2  &  3  &   60 &   1.1      \\
  1  &  1  &  0  &  7/2  &  3  &   60 &   1.2      \\ \hline
\end{tabular}
\label{tab7}
\end{table}

\begin{table}
\caption{The same as table \ref{tab3} for the 3$^-_2$ state in $^{18}$Ne
with the 0$^+$ core ground state using potential I. }
\begin{tabular}{|cccccc|cc|}
\hline
$\ell_x$ & $\ell_y$ & $L$ & $s_x$ & $S$ & $K_{max}$ & $W_{I}$ &
$W_{III}$  \\ \hline
 0  &  3  &  3  &  0  &  0  &  120 &  54.8  & 55.7   \\
 1  &  2  &  2  &  1  &  1  &  100 &   31.0  & 31.3   \\
 1  &  2  &  3  &  1  &  1  &    60 &   1.0  &  0.9  \\
 2  &  1  &  3  &  0  &  0  &    80 &   11.1  & 10.0   \\
 3  &  0  &  3  &  1  &  1  &    60 &   2.1  &  2.1  \\ \hline
 1  &  2  &  2  &  1/2  &  1  &   80 &   15.6  & 15.7   \\
 1  &  2  &  3  &  1/2  &  0  &  100 &   27.9  & 27.9   \\
 1  &  2  &  3  &  1/2  &  1  &   60 &   2.2  &  2.1  \\
 2  &  1  &  2  &  1/2  &  1  &   80 &   17.4  & 17.4   \\
 2  &  1  &  3  &  1/2  &  0  &  100 &   33.3  & 33.1   \\
 2  &  1  &  3  &  1/2  &  1  &   60 &   2.3  & 2.2   \\
 0  &  3  &  3  &  1/2  &  0  &   60 &   0.3  & 0.3   \\
 3  &  0  &  3  &  1/2  &  0  &   60 &   0.9  & 1.1   \\ \hline
\end{tabular}
\label{tab7a}
\end{table}

\begin{table}
\caption{The same as table \ref{tab3} for the 2$^-$ states in $^{18}$Ne
with the 3$^{-}$ core excited state. }
\begin{tabular}{|cccccc|c|}
\hline
$\ell_x$ & $\ell_y$ & $L$ & $s_x$ & $S$ & $K_{max}$ & $W_{IV}$  \\ \hline
  1  &  1  &  2  &  1  &  2  &   80 &    19.8     \\
      &      &     &      &      &       &     75.2     \\
  1  &  1  &  2  &  1  &  3  &   80 &     9.6     \\
      &      &     &      &      &       &     18.9     \\
  1  &  1  &  2  &  1  &  4  &   80 &     2.6     \\
      &      &     &      &      &       &     0.6     \\
  2  &  0  &  2  &  0  &  3  &  100 &    24.8     \\
      &      &     &      &      &       &     2.4     \\
  0  &  2  &  2  &  0  &  3  &  100 &    43.2   \\
      &      &     &      &      &       &     2.9     \\ \hline
  2  &  0  &  2  &  5/2  &  2  &  100 &  11.5       \\
      &      &     &      &      &       &     34.8     \\
  2  &  0  &  2  &  5/2  &  3  &  100 &  25.8       \\
      &      &     &      &      &       &     7.4     \\
  2  &  0  &  2  &  7/2  &  3  &  100 &  11.7       \\
      &      &     &      &      &       &     7.9     \\
  2  &  0  &  2  &  7/2  &  4  &   80 &   1.5       \\
      &      &     &      &      &       &     0.7        \\
  0  &  2  &  2  &  5/2  &  2  &  100 &  11.4       \\
      &      &     &      &      &       &     39.0        \\
  0  &  2  &  2  &  5/2  &  3  &  100 &   7.6       \\
      &      &     &      &      &       &     8.9        \\
  0  &  2  &  2  &  7/2  &  3  &  100 &  29.0       \\
      &      &     &      &      &       &     4.7        \\
  0  &  2  &  2  &  7/2  &  4  &   80 &   1.5       \\
      &      &     &      &      &       &     0.6        \\ \hline
\end{tabular}
\label{tab10}
\end{table}


\end{document}